\documentclass[11pt,a4paper,epsf,epsfig,psfrag]{article}
\usepackage{jheppub}
\usepackage{amsmath,graphicx,amsfonts, mathrsfs,amssymb}
\usepackage{amsmath,epsfig}
\usepackage{amssymb,amsfonts}
\usepackage{color}
\usepackage{latexsym}
\usepackage{epsfig}
\usepackage{pdfsync}
\newbox\pippobox

\def\be{\begin{equation}}
\def\ee{\end{equation}}
\def\bea{\begin{eqnarray}}
\def\eea{\end{eqnarray}}

\newcommand{\beq}{\begin{equation}}
\newcommand{\eeq}{\end{equation}}
\newcommand{\beqa}{\begin{eqnarray}}
\newcommand{\eeqa}{\end{eqnarray}}
\newcommand{\beqar}{\begin{eqnarray*}}
\newcommand{\eeqar}{\end{eqnarray*}}

\renewcommand{\eqref}[1]{(\ref{#1})}

\catcode`\@=12

\title{Chiral Phase Transition in the Soft-Wall Model of AdS/QCD }
\author[a,b,c]{Kaddour Chelabi,}
\author[a,b]{Zhen Fang,}
\author[d,e]{Mei Huang,}
\author[a]{Danning Li,}
\author[a,b]{Yue-Liang Wu}

\affiliation[a]{State Key Laboratory of Theoretical Physics,
Institute of Theoretical Physics, Chinese Academy of Sciences,
Beijing 100190, P. R. China }
\affiliation[b]{University of Chinese Academy of Sciences (UCAS), P.R. China}
\affiliation[c]{Laboratory of Particle Physics and Statistical Physics, Ecole Normale Superieure-Kouba. B.P. 92,16050, Vieux-Kouba, Algiers, Algeria}
\affiliation[d]{Institute of High
Energy Physics, Chinese Academy of Sciences, Beijing 100049, P.R. China}
\affiliation[e]{Theoretical Physics Center for Science
Facilities, Chinese Academy of Sciences, Beijing 100049, P.R. China}

\abstract{We investigate the chiral phase transition in the soft-wall model of AdS/QCD at zero chemical potential for two-flavor and three-flavor cases, respectively. We show that there is no spontaneous chiral symmetry breaking in the original soft-wall model. After detailed analysis, we find
that in order to realize chiral symmetry breaking and restoration, both profiles for the scalar potential and the dilaton field are essential. The scalar potential determines the possible solution structure of the chiral condensate, except the mass term, it takes another quartic term for the two-flavor case, and for the three-flavor case, one has to take into account an extra cubic term due to the t'Hooft determinant interaction. The profile of the dilaton field reflects the gluodynamics, which is negative at a certain ultraviolet scale and approaches positive quadratic behavior at far infrared region. With this set-up, the spontaneous chiral symmetry breaking in the vacuum and its restoration at finite temperature can be realized perfectly. In the two-flavor case, it gives a second order chiral phase transition in the chiral limit, while the transition turns to be a crossover for any finite quark mass. In the case of three-flavor, the phase transition becomes a first order one in the chiral limit, while above sufficient large quark mass it turns to be a crossover again. This scenario agrees exactly with the current understanding on chiral phase transition from lattice QCD and other effective model studies.}


\begin{document}
\maketitle

\section{Introduction}
\label{sec-int}
The vacuum of Quantum Chromodynamics(QCD), which is well known as the theory of Strong Interactions, is characterized by spontaneous chiral symmetry breaking and color charge confinement. And it is widely believed that at sufficient high temperature and/or density phase transition would happen in the system: chiral symmetry could be restored and color degrees of freedom can be freed. At present, to understand the phase structure of these two phase transitions is an important topic in both non-perturbative QCD study and cosmology\cite{nature-PTD}.

The properties of QCD phase transition depend sensitively on the inertial quantities of the system, such as the number($N_f$) of flavors and the mass of quarks($m_u,m_d$ and $m_s$). The chiral phase transition and confinement/deconfinement phase transition are well defined as a true phase transition only in the chiral limit and in the infinite quark mass limit respectively, when the chiral symmetry and $Z_3$ centre symmetry are the exact symmetries of QCD. In these cases, chiral condensate $\langle\bar{\psi}\psi\rangle$ and Polyakov loop $\langle L \rangle$ are well defined order parameters for chiral restoration and color charge deconfinement respectively. In the physical quark mass region, there are no known exact symmetries and the phase transitions are widely accepted as a continual transition, called crossover.

\begin{figure}[h]
\begin{center}
\epsfxsize=7.5 cm \epsfysize=7.5 cm \epsfbox{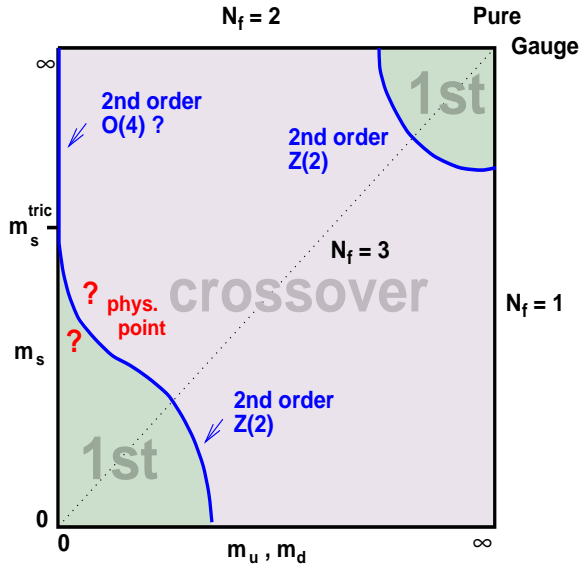} \
\end{center}
\caption{The expected phase diagram in the quark mass $m_u=m_d, m_s$ space(Taken from \cite{qcd-phase-diagram}).} \label{columbia-plot}
\end{figure}

Based on theoretical consideration and lattice QCD simulation\cite{qcd-phase-diagram,deForcrand:2006pv,Kanaya:2010qd}, the expected three flavor phase diagram in the quark mass plane can be summarized in the sketch plot (sometimes called "Colombia Plot") shown in Fig.\ref{columbia-plot} (Taken from \cite{qcd-phase-diagram}). Both near vanishing current quark mass region $m_u=m_d=m_s=0$ and near the infinite current quark mass region $m_u=m_d=m_s=\infty$, the phase transitions are widely accepted to be of first order, while in the intermediate region it becomes crossover. The boundaries between different regions are second order lines. In this paper, we will first focus on the two light flavor region with $m_u=m_d\simeq O(\text{MeV})$ and $m_s=\infty$ and the $SU(3)$ diagonal line with $m_u=m_d=m_s$. For the $SU(2)$ case the phase transition starts from a second order one in the chiral limit $m_u=m_d=0$ (in analogy to $O(4) ~ \sigma$ model\cite{Pisarski:1983ms}noting that $SU(2)_L\times SU(2)_R\simeq O(4)$). Then it is expected that even very small quark mass would drive the second order transition to a crossover one, as shown in Fig.\ref{columbia-plot}. For the $SU(3)$ case, the phase transition would start from a first order one in chiral limit and turn to crossover one at sufficient large quark mass.

The dominant physics for QCD phase transition is non-perturbative, hence perturbative methods become invalid in this region. Lattice QCD simulations are widely accepted as the most reliable method to study non-perturbative properties of QCD. However, despite the significant improvements which have been done in lattice QCD studies, it is still unable to get full understanding on QCD phase diagram from this ab initial approach. For example, the serious difficulty called sign problem prevents direct lattice simulations of QCD at finite chemical potential. Therefore, it is quite necessary to develop other non-perturbative methods to study the non-perturbative dynamics of QCD. In recent decades, the discovery of the anti-de Sitter/conformal field theory (AdS/CFT) correspondence and the conjecture of the gravity/gauge duality \cite{Maldacena:1997re,Gubser:1998bc,Witten:1998qj} provides a new powerful tool to solve the strong coupling problem of gauge theory, and shed lights on the full understanding of QCD phase transitions.

By breaking the conformal symmetry in different ways, many efforts have been made towards more realistic holographical description of QCD in non-perturbative region, such as hadron physics \cite{Erlich:2005qh,Karch:2006pv,TB:05,DaRold2005,D3-D7,D4-D6,SS-1,SS-2,Csaki:2006ji,Dp-Dq,Gherghetta-Kapusta-Kelley,Gherghetta-Kapusta-Kelley-2,YLWu,YLWu-1,Cui:2013xva,Li:2012ay,Li:2013oda,Colangelo:2008us} and hot/dense QCD matter \cite{Shuryak:2004cy,Tannenbaum:2006ch,Policastro:2001yc,Cai:2009zv,Cai:2008ph,Sin:2004yx,Shuryak:2005ia,Nastase:2005rp,
Janik:2005zt,Nakamura:2006ih,Sin:2006pv,Herzog:2006gh,Gubser-drag,Wu:2014gla,Li:2014dsa,Li:2014hja}, both in top-down approaches and in bottom-up approaches (see \cite{Aharony:1999ti,Erdmenger:2007cm,deTeramond:2012rt,Kim:2012ey,Adams:2012th} for reviews). For QCD phase transitions, most of the bottom-up studies \cite{Herzog:2006ra,BallonBayona:2007vp,Kim:2007em,Kim:2009wt,Cai:2007zw,Cai:2012xh,Gubser:2008ny,Gubser:2008yx,Gursoy:2008bu,Gursoy:2008za,Andreev-T3,Colangelo:2010pe,Li:2011hp,Cai:2012eh,Yaresko:2013tia,Yaresko:2015ysa,Noronha:2010mt,Finazzo:2014zga,He:2013qq,Yang:2014bqa,Yang:2015aia,Cui:2014oba,Afonin:2014jha,Zuo:2014iza,Zuo:2014vga} focus only on confinement/deconfinement phase transition. In these studies, in some sense, the geometric phases are distinguished by whether the expectation value of Polyakov Loop $\langle \text{L}\rangle$ calculated in the dual gravity background is vanishing or not. However, lattice simulations \cite{Aoki:2006br,Aoki:2009sc} have shown that the extracted phase transition temperatures from chiral condensate $\langle \bar{\psi}\psi\rangle$ and Polyakov Loop $\langle \text{L}\rangle$ are different for physical quark mass. For chiral restoration in $u,d$ sectors, the transition temperature is around $T_c^{\chi,(u,d)}=151 \text{MeV}$ while for confinement/deconfinement the transition temperature is around $T_c^d=176 \text{MeV}$. Therefore, it is necessary to study the temperature dependent chiral condensate and try to get the information of chiral phase transition in holographic framework.

In bottom-up approaches, hard-wall model \cite{Erlich:2005qh} and soft-wall model \cite{Karch:2006pv} are successful in describing hadron physics. Their extended models \cite{Gherghetta-Kapusta-Kelley,Gherghetta-Kapusta-Kelley-2,YLWu,YLWu-1,Cui:2013xva,Li:2012ay,Li:2013oda,Colangelo:2008us} can describe the hadron spectra and the related quantities in very good accuracy. In these models, the chiral condensate is introduced to realize chiral symmetry breaking at zero temperature. However, unlike in Nambu-Jona-Lasinio(NJL) model \cite{Nambu:1961tp,Nambu:1961fr}, the value of the chiral condensate is not self-consistently solved from the models themselves. Instead, it is often taken as a free parameter to fit the hadron spectra at zero temperature. The authors of \cite{Colangelo:2011sr} note that the IR boundary condition may require the dependence of chiral condensate on quark mass and try to solve chiral condensate self-consistently. By introducing black hole background and $U(1)$ gauge field, they extended the calculation to finite temperature and density, in such region they get the phase diagram for chiral phase transition which agrees with the previous approaches(see e.g.\cite{Barducci:1994}). However, since their studies based on the simplest assumption on the gravity background as the original soft-wall model, the quark mass dependence of the chiral condensate and chiral phase transition is not analysed in their work which is our main case in our paper. Noting that the dilaton profile is in the central place to generate correct quark mass dependence behavior of chiral condensate, we try to study the effects of dilaton profile on the chiral phase transition in \cite{Chelabi:2015cwn}. This work is an extension of \cite{Chelabi:2015cwn}, and we will study the dilaton and scalar potential effects in details.

The paper is organized as follows. In Sec.\ref{soft-ori}, we explain why in the previous set-up the quark mass dependence behavior is not correct, and why a quartic potential term of the scalar field is necessary. Then in Sec.\ref{sec-quartic}, we study the effects of scalar potential and dilaton profile on chiral condensate, and we show that under the negative quadratic dilaton model and quartic scalar potential the quark-mass dependence behavior of chiral phase transition is realized correctly, though there is massless scalar meson state in this model. In Sec.\ref{sec-interpolation}, to get rid of the massless scalar mode, we propose a dilaton profile negative at certain ultraviolet scale and positive at far infrared scale, and we show that chiral phase transition could be well described in this model. In Sec.\ref{sum}, a short summary and discussion are listed. We also describe the numerical process we used to extract chiral condensate in Appendix.\ref{appendix-sec1}.

\section{Chiral symmetry breaking in original soft-wall model}
\label{soft-ori}

As we mentioned in the above, soft-wall model provides a start point to study linear confinement and chiral symmetry breaking of QCD in bottom-up approach, and the prediction of meson spectrum in its extended models is in good agreement with the experimental data\cite{Gherghetta-Kapusta-Kelley,Gherghetta-Kapusta-Kelley-2,YLWu,YLWu-1,Cui:2013xva,Li:2012ay,Li:2013oda}. Therefore it is interesting to investigate the spontaneous chiral symmetry breaking of QCD and its restoration in this model. In this section we will give a brief review of this model and then try to study the temperature dependent behavior of chiral condensate, which is the order parameter of chiral phase transition. It should be noted that the similar study in this section has been studied in \cite{Colangelo:2011sr}. However there the authors have not studied the quark mass dependence of chiral condensate, which is of the main interest of this manuscript.

\subsection{General setup}
\label{sec-setup}

In the original paper of soft-wall model\cite{Karch:2006pv}, the authors promote the 4D global chiral symmetry $SU(2)_L\times SU(2)_R$ of QCD to 5D, and consider the following action
\begin{eqnarray}\label{kkssaction}
 S=&&-\int d^5x
 \sqrt{-g}e^{-\Phi}Tr(D_m X^+ D^m X+M_5^2 X^{+}X+\frac{1}{4g_5^2}(F_L^2+F_R^2)),
\end{eqnarray}
with $A_{L/R}$ the left/right hand gauge field, $D_m$ the covariant derivative defined as $D_mX=\partial_mX-i A^L_mX+i XA^R_m$, $F_{mn}$the field strength defined as $F_{mn}=\partial_m A_{n}-\partial_n A_{m}-i[A_m,A_n]$, $g$ the determinant of metric, and $\Phi$ the dilaton field. The mass of the complex scalar field $X$ $M_5^2$ can be determined as $M_5^2=-3$(we take the AdS radius $L=1$ in this work) from the AdS/CFT prescription $M_5^2=(\Delta-p)(\Delta+p-4)$\cite{Witten:1998qj} by taking $\Delta=3, p=0$.

If the scalar field $X$ gets a non-vanishing vacuum expectation value $X_0$, then the $SU(2)_L\times SU(2)_R$ symmetry would be broken, in which the soft-wall model can mimic chiral symmetry breaking of QCD. If one assumes that $m_u=m_d$, the symmetry should be broken to $SU(2)$ and $X_0$ would have the form $\frac{\chi(z)}{2}I_2$. Here $I_2$ is the $2\times 2$ identity matrix and $\chi(z)$ is assumed to depend only on the fifth coordinate $z$. It should be noted that by considering $X$ as an $N_f\times N_f$ matrix and taking the generators from $SU(N_f)$, it is easy to extend this model to $SU(N_f)_L\times SU(N_f)_R$ case. The same procedure can be done by replacing $X_0=\frac{\chi(z)}{\sqrt{2N_f}}I_{N_f}$. Here the factor $\sqrt{2N_f}$ is chosen to keep the kinetic term of $\chi$ canonical. After these assumptions, the effective description of the vacuum expectation value of $X$ in terms of $\chi$ reads
\begin{eqnarray}\label{action-chi-0}
S_{\chi}=-\int d^5x
 \sqrt{-g}e^{-\Phi}(\frac{1}{2}g^{zz}\chi^{'2}+\frac{1}{2}M_5^2 \chi^2),
\end{eqnarray}
where $'$ denotes the derivative with respect to $z$.

Under the metric ansatz
\begin{eqnarray}\label{metric-T0}
ds^2=e^{2A_s(z)}(-dt^2+dz^2+dx_idx^i),
\end{eqnarray}
$\chi(z)$ could be solved from the following equation
\begin{eqnarray}\label{eom-chi}
\chi^{''}+(3A_s^{'}-\Phi^{'})\chi^{'}+3e^{2A_s}\chi=0.
\end{eqnarray}
From this equation, the ultraviolet(UV,$z\rightarrow 0$) asymptotic behavior could be solved as
\begin{eqnarray}\label{chi-UV}
\chi(z)=c_1 z+c_3 z^3+...,
\end{eqnarray}
with $c_1,c_3$ two integral constants of the second order ordinary derivative equation(ODE). From the assumption, $X$ is dual to $\bar{q}q$ operator, so one can related $c_1, c_3$ to the quark mass and chiral condensate as

\begin{eqnarray}\label{c1c3}
c_1=m_q \zeta, c_3=\sigma/\zeta
\end{eqnarray}
with the normalization factor $\zeta=\frac{\sqrt{N_c}}{2\pi}$ \cite{Cherman:2008eh}.Then, the meson spectral in this model can be read from poles in corresponding Green functions, or equivalent by solving the following Schrodinger-like equations(For details, see \cite{Karch:2006pv}, and also \cite{Li:2012ay,Li:2013oda})

\begin{eqnarray}\label{scalar-sn}
& & -s_n^{''}+V_s(z)s_n=m_n^2s_n,  \\
& & -\pi_n''+V_{\pi,\varphi}\pi_n=m_n^2(\pi_n-e^{A_s}\chi\varphi_n),  \\
& &  -\varphi_n''+ V_{\varphi} \varphi_n=g_5^2 e^{A_s}\chi(\pi_n-e^{A_s}\chi\varphi_n), \\
& & -v_n^{''}+V_v(z)v_n=m_{n,v}^2v_n, \label{vector-n} \\
& & -a_n^{''}+V_a a_n = m_n^2 a_n,
\end{eqnarray}
with the schrodinger-like potentials
\begin{eqnarray}
&&V_s=\frac{3A_s^{''}-\phi^{''}}{2}+\frac{(3A_s^{'}-\phi^{'})^2}{4}+3e^{2A_s},
\label{s-vz}\\
&& V_{\pi,\varphi}=\frac{3A_s^{''}-\phi^{''}+2\chi^{''}/\chi-2\chi^{'2}/\chi^2}{2} \label{ps-vz} \nonumber \\
 & & ~~~~~~~~~~ +\frac{(3A_s^{'}-\phi^{'}+2\chi^{'}/\chi)^2}{4},  \\
&& V_{\varphi} = \frac{A_s^{''}-\phi^{''}}{2}+\frac{(A_s^{'}-\phi^{'})^2}{4},  \\
&& V_v=\frac{A_s^{''}-\phi^{''}}{2}+\frac{(A_s^{'}-\phi^{'})^2}{4} ~, \label{v-vz} \\
&& V_a = \frac{A_s^{'}-\phi^{'}}{2}+\frac{(A_s^{'}-\phi^{'})^2}{4}+g_5^2 e^{2A_s}\chi^{2}
\label{a-vz}.
\end{eqnarray}
Here, $s_n,\pi_n,v_n,a_n$ are the 5D wave function of scalar, pseudoscalar, vector and axial-vector meson respectively.

In the original model, the authors took the $AdS_5$ metric, i.e. $A_s=-\log(z)$ in Eq.(\ref{metric-T0}), and one can check that if $\Phi(z)$ increases as $z^2$ in Infrared(IR) region($z\rightarrow \infty$), the mass square of the highly excited meson states would be proportional to the radial quantum number $n$, i.e. $m_n^2\propto n$ when $n>> 1$. Therefore, in the original model, the authors simply took $\Phi\propto z^2$ to realize the linear spectral. Furthermore, one can read from Eqs.(\ref{v-vz},\ref{a-vz}) that if $\chi\neq0$, the equations of motion are different for vector and axial-vector mesons, which shows the possibility to realize chiral symmetry breaking in this model.

In this section, we have seen that the quadratic dilaton at IR and the non-vanishing scalar $\chi$ are important for linear confinement and chiral symmetry breaking of QCD vacuum respectively. The linear confinement has already been studied widely in the extended models of the original soft-wall model, and it turns out that soft-wall model does give success description in this topic. However, the chiral symmetry breaking and its restoration are much less studied in this framework. Therefore, in the next section we will focus on this topic.

\subsection{Quark mass and temperature dependence of chiral condensate}
\label{sec-ori-positive}

In this section, we would like to study the quark mass and temperature dependence of chiral condensate in the original soft-wall model. The main purpose is to check whether the symmetry breaking is a spontaneous one as that in real QCD or not.

We start from the soft-wall action Eq.(\ref{action-chi-0}), and under the usual metric ansatz at finite temperature
\begin{eqnarray}\label{metric-ansatz}
d s^2=e^{2A_s(z)}(-f(z) d t^2+\frac{1}{f(z)}d z^2+dx_i dx^i),
\end{eqnarray}
the equation of motion for $\chi(z)$ can be derived as

\begin{eqnarray}\label{eom-chi-0}
\chi^{''}+(3A_s^{'}-\Phi^{'}+\frac{f^{'}}{f})\chi^{'}+3 \frac{e^{2A_s}}{f}\chi=0.
\end{eqnarray}

As in the original soft-wall model, we take the dilaton field of the simple quadratic form
\begin{eqnarray}\label{dilaton-sol}
\Phi(z)=\mu^2 z^2,
\end{eqnarray}

and the background metric as the AdS-Schwarzchild black hole solution

\begin{eqnarray}
A_s(z)&=&-\log(z),\label{adsas}\\
f(z)&=&1-\frac{z^4}{z_h^4}.\label{adsf}
\end{eqnarray}
Here $z_h$ is the horizon of the black hole defined at $f(z_h)=0$ and related to the temperature $T$ of the system by the Hawking temperature formula

\begin{eqnarray}
T=|\frac{f^{'}(z_h)}{4\pi}|=\frac{1}{\pi z_h}.
\end{eqnarray}

\begin{figure}[h]
\begin{center}
\epsfxsize=6.5 cm \epsfysize=6.5 cm \epsfbox{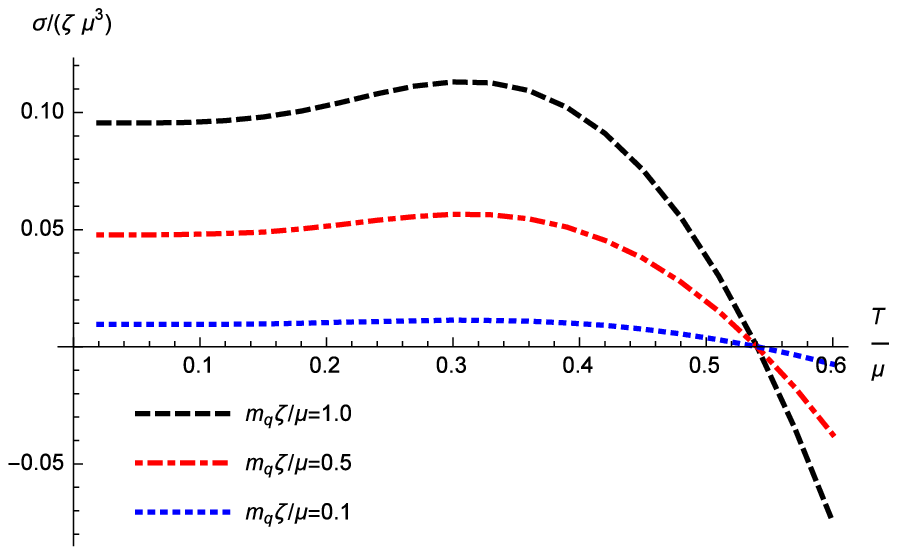}
\hspace*{0.1cm} \epsfxsize=6.5 cm \epsfysize=6.5 cm
\epsfbox{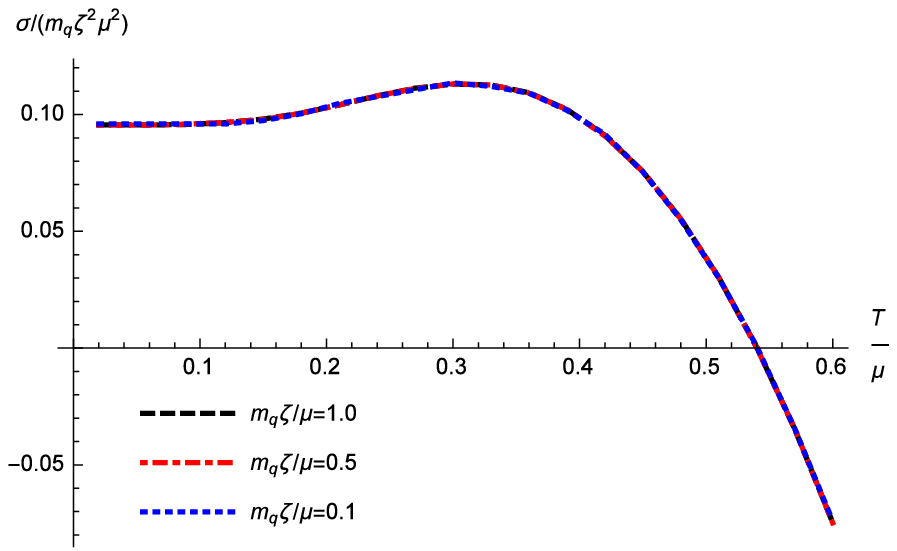} \vskip -0.05cm \hskip 0.15 cm
\textbf{( a ) } \hskip 6.5 cm \textbf{( b )} \\
\end{center}
\caption{$\sigma(T)$ as a function of temperature $T$ in quadratic dilaton background Eq.(\ref{dilaton-sol}). Panel (a) gives the results of $\sigma/(\zeta\mu^3)$. The blue dotted, red dotdashed,and black dashed lines represent $m_q \zeta/\mu=0.1,0.5,1$ respectively . Panel (b) gives the $m_q$ rescaled dimensionless results $\sigma/(m_q\zeta^2\mu^2)$ of the three lines. All the three lines overlap and become the same curve, which shows that $\sigma/m_q$ does not depend on $m_q$.}
 \label{kkss-sigma}
\end{figure}

At zero temperature, we have $f(z)\equiv1$, Eq.(\ref{eom-chi-0}) could be analytically solved as
\begin{eqnarray}\label{chi-sol-ori}
\chi(z)=c_2 G_{1,2}^{2,0}\left(-\mu^2z^2|
\begin{array}{c}
 1 \\
 \frac{1}{2},\frac{3}{2}
\end{array}
\right)+c_1 e^{\frac{\mu^2z^2}{2}} \mu^3z^3 \left(I_0\left(\frac{\mu^2z^2}{2}\right)
+I_1\left(\frac{\mu^2z^2}{2}\right)\right).
\end{eqnarray}
with $I_n(z)$ the modified Bessel function of the first kind and $G_{pq}^{mn}\left(z\left|
\begin{array}{c}
 a_1,\ldots ,a_p \\
 b_1,\ldots ,b_q
\end{array}
\right.\right)$ the MeijerG function \footnote{For details, please refer to Mathematica 10}. The $z\rightarrow \infty$ asymptotic behavior of Eq.(\ref{chi-sol-ori}) is either exponentially blowing up or approaching a constant, which can be seen by taking two limits : $\chi^{''}<<\chi$, then $-2\mu^2 z \chi^{'}+3\chi/z^2=0$ and $\chi_1 \rightarrow e^{-3/(4 \mu^2 z^2)}\rightarrow 1$; $\chi^{'}>>\chi$, then $\chi^{''}-2\mu^2 z \chi^{'}=0$ and $\chi_2 \rightarrow e^{\mu^2 z^2}/(\mu z)$. The exponentially growing branch is unacceptable, since it breaks the Regge behavior of axial vector in soft-wall model. Then because both the two independent solutions in Eq.(\ref{chi-sol-ori}) blow up in the IR region, we will require $c_1\propto c_2$ to cancel the exponential growing part. Finally, we can get the ultraviolet(UV) $z\rightarrow 0$ asymptotic behavior of the solution of Eq.(\ref{chi-sol-ori}) as
\begin{eqnarray}\label{kkss-chi}
\chi(z)&&=c(\mu z+\mu^3z^3 \left(-\frac{1}{2}
+\gamma_E +\frac{\psi \left(-\frac{1}{2}\right)}{2}+\log (\mu z)\right))+O(z^4)\\
&&=c(\mu z+\mu^3z^3(0.095+\log(\mu z)))
\end{eqnarray}
with $\gamma_E=0.577$ the Euler's constant and $\psi(z)$ the digamma function with $\psi(-\frac{1}{2})=0.036$. Comparing Eq.(\ref{c1c3}) and Eq.(\ref{kkss-chi}), one gets $c=m_q\zeta/\mu$ and
\begin{eqnarray}
\frac{\sigma(T=0)}{m_q \mu^2\zeta^2}=0.095 .
\end{eqnarray}
From this result, we can see that the chiral condensate in the original soft-wall model is induced by the quark mass, and in the chiral limit $m_q\rightarrow0$ it tends to vanish. In this sense, the original soft-wall model can not describe the spontaneous chiral symmetry breaking of QCD vacuum. The symmetry breaking is actually an explicit one.

At finite temperature, Eq.(\ref{eom-chi-0}) can not be solved analytically. By requiring $\chi(z)$ taking the asymptotic behavior as Eq.(\ref{chi-UV}) at UV  and all the quantities being regular at horizon $z_h$, the authors of Ref.\cite{Colangelo:2011sr} get the numerical result as shown in the dashed line in Fig.\ref{kkss-sigma}(a) in case $m_q=\mu$. In the plot, we scaled all the quantities with dimension to dimensionless quantities using the only mass scale $\mu$ in the model. In Fig.\ref{kkss-sigma}(a), we also show the $m_q$ dependence of $\sigma(T)$ explicitly. From Fig.\ref{kkss-sigma}(b) we see that the $\sigma/m_q$ results do not depend on $m_q$, which shows that $\sigma(T)= m_q g(T)$ with $g(T)$ being independent of $m_q$. One can also find that the near $T=0$ value of $\sigma/(m_q\zeta^2\mu^2)$ is just the same value $0.095$ as extracted analytically, which can be seen as a consistent check of the numerical method we used(We leave this part in Appendix.\ref{appendix-sec1}).

In fact, it is also easy to show this conclusion analytically: since Eq.(\ref{eom-chi-0}) is linear, the solution would be of the general form $\chi(z)_T=c_1(T) g_1(z)+c_2(T) g_2(z)$, where $g_1(z)$ and $g_2(z)$ is singular at $z_h$. In order to cancel the singularity at $z_h$, $c_1(T)$ should be proportional to $c_2(T)$. In addition, since $m_q$ is fixed at any temperature, we have $\sigma(T)\propto c_1(T)\propto m_q$. Therefore in the original soft-wall model, in the chiral limit $m_q\rightarrow0$, the chiral condensate would vanish. So at finite temperature, we can also see that chiral symmetry breaking in the original soft-wall model is an explicit one.

In summary, the chiral condensate in the original model is induced by the quark mass, which is unreasonable from real QCD point of view. In addition, from the analysis, we could see that the main problem in the original model is that Eq.(\ref{eom-chi-0}) is a linear equation. Therefore, in the next section we will add non-linear terms in the potential of scalar field $X$ and try to solve this problem.

\section{Quartic potential effects}
\label{sec-quartic}

In the previous section, we have seen that the model in Sec.\ref{sec-ori-positive} can not give the correct chiral symmetry breaking mechanism at both zero and finite temperature. In order to modify the linear quark mass dependence behavior of chiral condensate $\sigma(T)$, it is necessary to add non-linear term in the potential of scalar field $X$, which has also been pointed out in \cite{Karch:2006pv} and has been studied in \cite{YLWu,YLWu-1,Cui:2013xva,Li:2012ay,Li:2013oda,Gherghetta-Kapusta-Kelley,Gherghetta-Kapusta-Kelley-2}. Here, we assume that the potential of scalar field $X$ takes a general form $V_X(|X|)$ in terms of $|X|$ and the action becomes

\begin{eqnarray}\label{action}
 S=&&-\int d^5x
 \sqrt{-g}e^{-\Phi}Tr(D_m X^+ D^m X+V_X(|X|)).
\end{eqnarray}

\begin{figure}[h]
\begin{center}
\epsfxsize=9.5 cm \epsfysize=7.5 cm \epsfbox{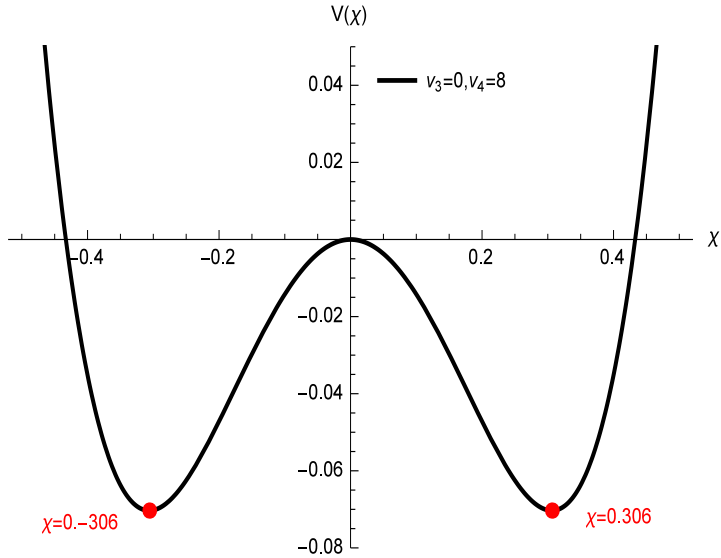} \
\end{center}
\caption{$V(\chi)$ as a function of $\chi$ with $v_3=0,v_4=8$. The two non-trivial solutions $\chi=\pm 0.306$ of $\partial_\chi V(\chi)=0$ are labeled.} \label{v4=8-potential}
\end{figure}

Inserting the expectation value of $X$, we get the the effective description in terms of $\chi$ of the following form

\begin{eqnarray}\label{eff-action}
S_{\chi}=-\int d^5x
 \sqrt{-g}e^{-\Phi}(\frac{1}{2}g^{zz}\chi^{'2}+V(\chi)),
\end{eqnarray}
where $V(\chi)\equiv Tr({V_X(|X|)})$. The leading term of $V(\chi)$ comes from the mass term and it is fixed to be $-\frac{3}{2}\chi^2$ from AdS/CFT dictionary. To keep $\chi\leftrightarrow-\chi$ symmetry, the next power term is the quartic term $v_4 \chi^4$. If one considers the $N_f=3$ case, there would be $\chi^3$ term coming from the t'Hooft determinant term $\text{Re}[det(X)]$. For later convenience, here we will consider

\begin{eqnarray}\label{vchi}
V(\chi)=-\frac{3}{2}\chi^2+v_3\chi^3+v_4\chi^4.
\end{eqnarray}
In this section, we will only consider the $N_f=2$ case and only study the corrections from the quartic term when $v_3=0, v_4\neq0$. Since in \cite{Gherghetta-Kapusta-Kelley,Gherghetta-Kapusta-Kelley-2,YLWu,YLWu-1}, $v_4\simeq10$ gives a better prediction of meson spectral, in our work we will choose the parameter $v_4$ in this region. In Fig.\ref{v4=8-potential}, we plot the potential of $v_4=8$, and we see that as in spontaneously broken of $\phi^4$ theory, $\partial_\chi V(\chi)=0$ has three solutions: $\chi=0$ and $\chi=\pm \sqrt{\frac{3}{4 v_4}}\simeq \pm 0.306$. We would see later that this structure plays an important role to realize the spontaneous symmetry breaking in 5D.

The equation of motion for $\chi$ with respect to the action Eq.(\ref{eff-action}) can be derived as

\begin{eqnarray}\label{eom-chi-1}
\chi^{''}+(3A_s^{'}-\Phi^{'}+\frac{f^{'}}{f})\chi^{'}- \frac{e^{2A_s}}{f}\partial_\chi V(\chi)=0.
\end{eqnarray}
and we checked that the leading UV expansion of this equation is still of the form $m_q \zeta z+\frac{\sigma}{\zeta} z^3$ \cite{Cherman:2008eh}.

For later convenience in comparing the stability of different solutions, we would like to derive the free energy of the solutions here. One can use the equivalence of the partition function conjecture $Z_{QCD}=Z_{gravity}\simeq e^{-S_E}$. Using the thermodynamical equality $Z=e^{-\beta F}$ ($\beta\equiv 1/T$), we have the relation between free energy $F$ and the on-shell Euclidean action of gravity $F=S_E$. Inserting Eq.(\ref{eom-chi-1}) into Eq.(\ref{eff-action}), and notice the negative sign while doing the wick rotation, we have

\begin{eqnarray}\label{freeenergy}
\mathcal{F}\equiv \frac{F}{V_3}&=&\int_0^{z_h} d z
 \sqrt{-g}e^{-\Phi}(\frac{1}{2}g^{zz}\chi^{'2}+V(\chi))\nonumber\\
 &=&\int_0^{z_h} dz \sqrt{-g}e^{-\Phi}  (\frac{1}{2}v_3\chi^3+v_4\chi^4)+\frac{1}{2}(\chi e^{3A_s-\Phi}f\chi^{'})|_{\epsilon}^{z_h}\nonumber\\
 &=&-\int_0^{z_h} dz  e^{5A_s-\Phi}  (\frac{1}{2}v_3\chi^3+v_4\chi^4)-\frac{1}{2}(\chi e^{3A_s-\Phi}f\chi^{'})|_{\epsilon},
\end{eqnarray}
where we have introduced the free energy density $\mathcal{F}$ to get rid of the infinite spatial volume $V_3$ in the integral of $S_E$.

When $m_q=0$, from the UV asymptotic behavior of $\chi$, it is easy to see that the last equation is regular at $\epsilon=0$ and the second term vanishes. In this case, $\chi\equiv0$ is always a solution of Eq.(\ref{eom-chi-1}), since it satisfies $m_q=0$ and is regular everywhere. This solution stands for the chiral symmetry restored phase, for $\sigma=0$. When $N_f=2, v_3=0$, if $v_4>0$, we could easily see that $\mathcal{F}<0$ if $\chi\neq 0$, which means in chiral limit if there are non-trivial solutions of $\chi$, it is always thermodynamically more favored than the symmetry restored solution $\chi\equiv0$. When $m_q\neq0$, $\mathcal{F}$ is divergent near $\epsilon=0$, and in principle one has to add counter term to cancel the divergence. Fortunately, later we do not really need to deal with this case, so here we do not try to work out the explicit form of the counter terms. After all this preparation, in the following subsections, we will try to analyze the quark mass and temperature dependent behavior of $\sigma(T)$ under the quatric corrections in the potential.

\subsection{Chiral condensate in positive quadratic dilaton background}
\label{sec-quartic-positive}

\begin{figure}[h]
\begin{center}
\epsfxsize=6.5 cm \epsfysize=6.5 cm \epsfbox{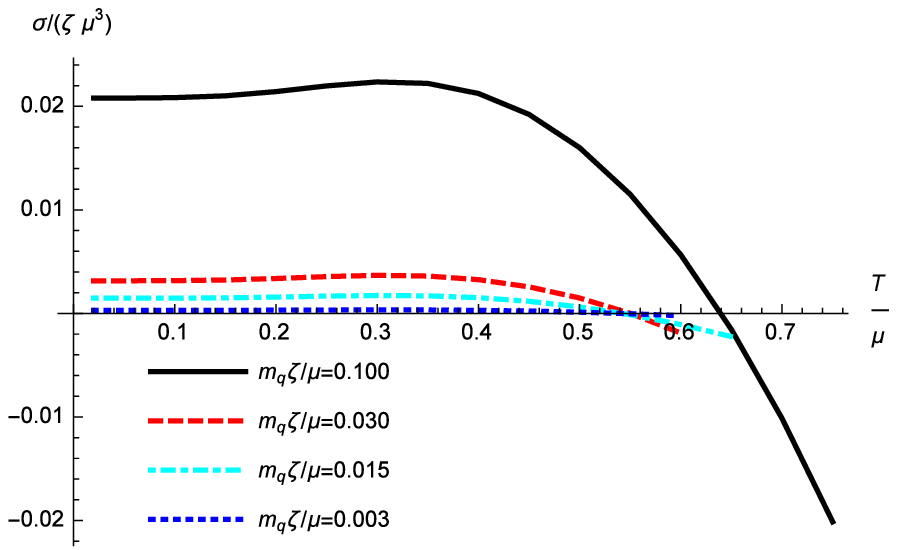}
\hspace*{0.1cm} \epsfxsize=6.5 cm \epsfysize=6.5 cm
\epsfbox{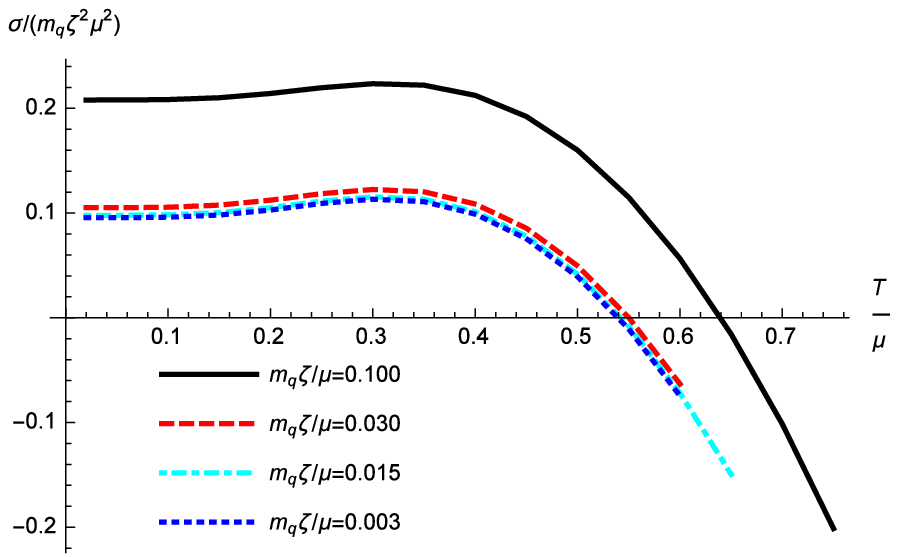} \vskip -0.05cm \hskip 0.15 cm
\textbf{( a ) } \hskip 6.5 cm \textbf{( b )} \\
\end{center}
\caption{$\sigma(T)$ as a function of temperature $T$ in quadratic dilaton background Eq.(\ref{dilaton-sol}) with $v_3=0,v_4=8$. Panel (a) gives the results of $\sigma/(\zeta \mu^3)$. The blue dotted, cyan dotdashed, red dashed and black solid lines represent $m_q \zeta/\mu=0.003,0.015,0.03, 0.1$ respectively . Panel (b) gives the $m_q$ rescaled dimensionless results $\sigma/(m_q\zeta^2\mu^3)$ of the three lines. When $m_q\rightarrow0$, the results of $\sigma/(m_q\zeta^2\mu^3)$ tend to be independent of $m_q$ and approach the lines in Fig.\ref{kkss-sigma}(b). } \label{sigma-T-m1-po-nl}
\end{figure}

Firstly, we work in the positive dilaton background as in the last section. We take $v_4=8$ and solve Eq.(\ref{eom-chi-1}) with the boundary condition $\chi(z\rightarrow0)\rightarrow m_q z, \chi(z\rightarrow z_h)<\infty$. We take $m_q \zeta/\mu=0.003,0.015,0.03, 0.1$ and show the results in Fig.\ref{sigma-T-m1-po-nl}. Since there is only one parameter with mass dimension, we plot the dimensionless combination $\sigma/(\zeta\mu^3)$ and $\sigma/(m_q\zeta^2\mu^2)$ in terms of $T/\mu$. It is very clear to see that those plots will not depend on $\mu$.

From Panel.(a), we can see chiral condensate at low temperature decrease with the decreasing of quark mass. From the plots of $\sigma/(m_q\zeta^2\mu^2)$ in Panel.(b), we could see that $\sigma$ is no longer simply proportional to $m_q$ due to the non-linear potential. In the large $m_q$ region, the non-linear potential would affect the result of $\sigma$ significantly. It increases the value of $\sigma/(m_q\zeta^2\mu^2)$ from $0.095$ to around $0.2$ at $T\simeq0$ region. However, when $m_q$ approaching the chiral limit, it can be seen that $\sigma/(m_q\mu^2\zeta^2)$ will approach a limit, which is just the results without the quartic potential term. Therefore, we see that in the small $m_q$ region, $\sigma\propto m_q$ is approximately true, which again shows that no chiral condensate in chiral limit and the symmetry breaking is still an explicit one induced by quark mass.

From the above discussion, unlike the 4D field theory, it is insufficient to get spontaneously symmetry breaking with only the $\phi^4$ like potential. Modification on the dilaton field or the metric background is necessary.

\subsection{Chiral condensate in negative dilaton model}
\label{sec-quartic-negative}

The positive dilaton model does not give the correct spontaneous symmetry breaking behavior in chiral limit. In this section, as a test, we will work in the negative dilaton background
\begin{eqnarray}\label{negativedilaton}
\Phi(z)=-\mu^2z^2,
\end{eqnarray}
though as pointed out in \cite{Karch:2006pv,Li:2012ay,KKSS-2} it might cause an un-physical massless scalar meson state.

\begin{figure}[h]
\begin{center}
\epsfxsize=6.5 cm \epsfysize=6.5 cm \epsfbox{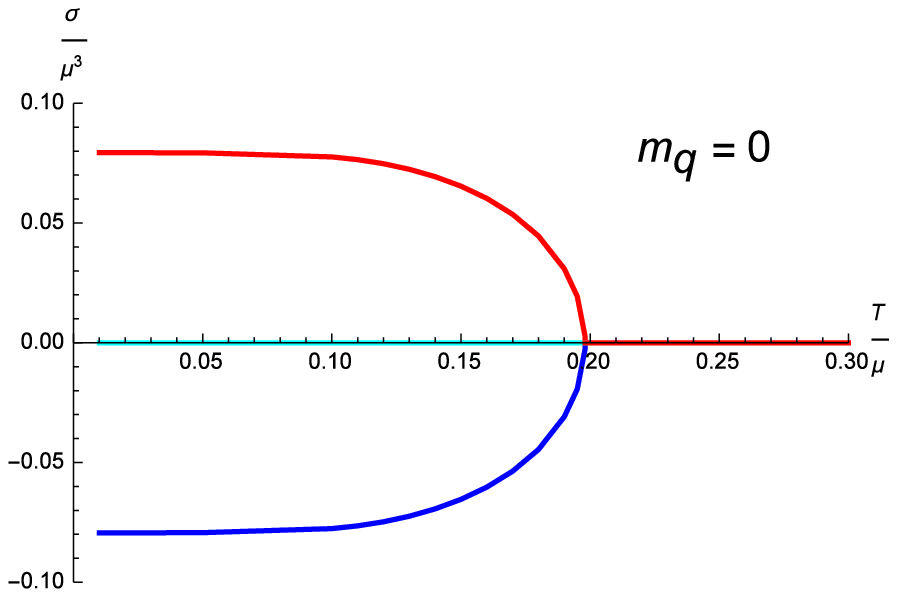}
\hspace*{0.1cm} \epsfxsize=6.5 cm \epsfysize=6.5 cm
\epsfbox{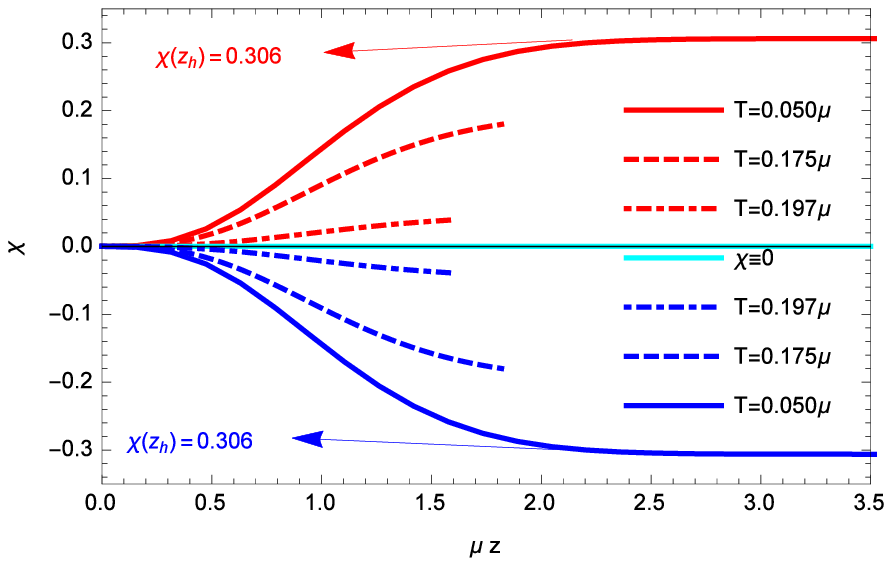} \vskip -0.05cm \hskip 0.15 cm
\textbf{( a ) } \hskip 6.5 cm \textbf{( b )} \\
\end{center}
\caption{Panel.(a) shows $\mu^3$ scaled dimensionless chiral condensate $\sigma(T)/\mu^3$ as a function of $\mu$ scaled temperature $T/\mu$ in negative quadratic dilaton background Eq.(\ref{negativedilaton}) when $v_3=0,v_4=8$ and $m_q=0$.  Panel.(b) shows the corresponding $\chi$ solutions when $T=0.050,0.175,0.197\mu$. We only plot the part outside the black hole $0<z<z_h=1/(\pi T)$ in each temperature and cut the long constant tail in the region $3.5<\mu z<20/\pi$ when $T=0.05\mu$ for compactness of the plots.}
\label{negative-sigma-m0}
\end{figure}

Firstly, we take $v_4=8$ and consider the chiral limit case. Taking $m_q=0$ in the UV expansion $\chi(z)=m_q \zeta+\sigma/\zeta+...$ and requiring $|\chi(z_h)|<\infty$ at horizon, we could solve $\sigma(T)$ from Eq.(\ref{eom-chi-1}) using the method described in Appendix.\ref{appendix-sec1}. The results are shown in Fig.\ref{negative-sigma-m0}, in which we have scaled all the quantities with dimension into dimensionless combination using $\mu$. From Fig.\ref{negative-sigma-m0}(a), we could see that below $T=0.198\mu$ there are two non-trivial $\sigma\neq0$ solutions labeled by red and blue solid lines together with the trivial $\chi\equiv0$ solution labeled by cyan solid line. In fact, the red and blue solutions are just the $\sigma\leftrightarrow-\sigma$ reflection of each other, which is the manifestation of the $\chi\leftrightarrow-\chi$ $Z_2$ symmetry of the scalar potential Eq.(\ref{vchi}) when $v_3=0$. Thermodynamically, the non-trivial solutions are more stable than the trivial $\sigma=0$ solution when $v_3=0$, as has been explained at the beginning of this section.  Furthermore, the non-trivial solutions of $\sigma$ approach a finite constant value $\sigma_0\equiv\sigma(T=0)\simeq0.08\mu^3=(0.43\mu)^3$ at low temperature, representing the spontaneous chiral symmetry breaking in the vacuum. Then, when $T$ increasing, $\sigma(T)$ decreases slowly from $0.08\mu^3$ to around $0.07\mu^3$ in the temperature region $0<T<0.140\mu$. Nextly, when the temperature continues to increase from $0.140\mu$ to $0.198\mu$, $\sigma(T)$ decreases rapidly from $0.07\mu$ to zero. Above $T=0.198\mu$, we can not find non-trivial solutions, which shows the restoration of chiral symmetry at high temperature. From this picture, we have realized the spontaneous chiral symmetry breaking in the vacuum $T=0$ and its restoration above the critical temperature $T^{negative}_C=0.198\mu$. The phase transition is a second order one in chiral limit, consistent with the `Columbia sketch plot' in two flavor chiral limit. We also note that if we take $\mu=0.75\rm{GeV}$, then vacuum value of chiral condensate and the critical temperature would be $\sigma_0\simeq(323\rm{MeV})^3$ and $T^{negative}_C\simeq150\rm{MeV}$, which is comparable with the lattice results \cite{Aoki:2006br,Aoki:2009sc}.

In order to show the effects of the two non-trivial vacua in scalar potential Eq.(\ref{vchi}) and Fig.\ref{v4=8-potential}, we also plot the corresponding solutions of $\chi(z)$ at temperature $T=0.05,0.175,0.197\mu$ in Fig.\ref{negative-sigma-m0}(b). From this figure, we could see that at low temperature $T=0.05\mu$, $\chi(z)$ would increase from $0$ to a finite constant value near the horizon. Actually, we find that even when $T$ is around $0.1\mu$, the configuration of $\chi(z)$ does not change too much. We extract the near horizon value and find that it equals to $0.306$, which is just the too non-trivial vacuum labeled in Fig.\ref{v4=8-potential}. Then when temperature increases, we see that the solution of $\chi(z)$ falls towards the $x$ axis, and when $T=0.197\mu$ $\chi(z)$ become very close to $\chi=0$. It is easy to imagine that when $T=T_C^{negative}$, the solution of $\chi(z)$ become exactly $\chi(z)=0$ at range of $0<\mu z<1/(\pi T)$.  Therefore, we see that the near $T=0$ solutions of $\chi(z)$ is just the interpolation between the trivial vacuum $\chi=0$ and the two non-trivial vacuum $\chi=\pm 0.306$, which in some sense shows the necessity of the positive quartic term in the scalar potential.

\begin{figure}[h]
\begin{center}
\epsfxsize=6.5 cm \epsfysize=6.5 cm \epsfbox{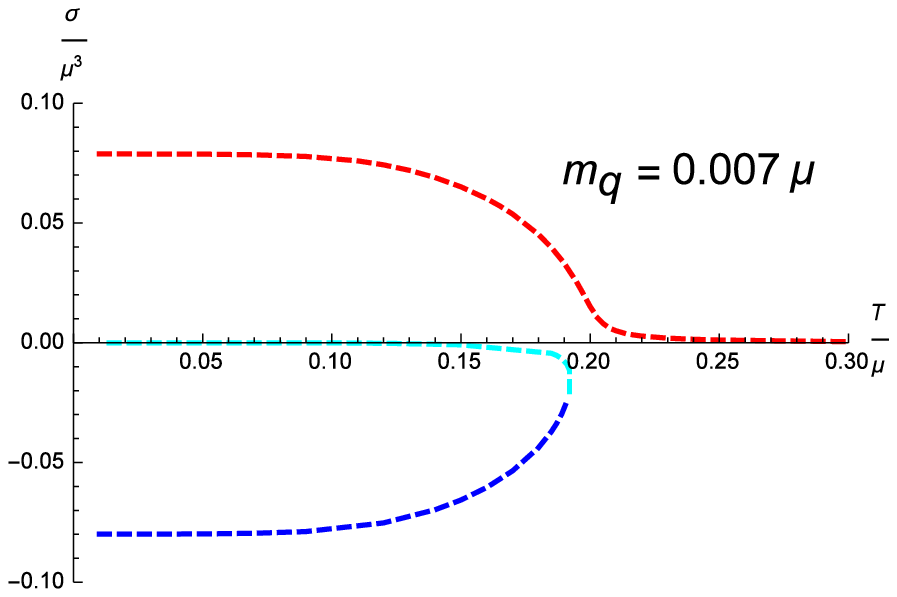}
\hspace*{0.1cm} \epsfxsize=6.5 cm \epsfysize=6.5 cm
\epsfbox{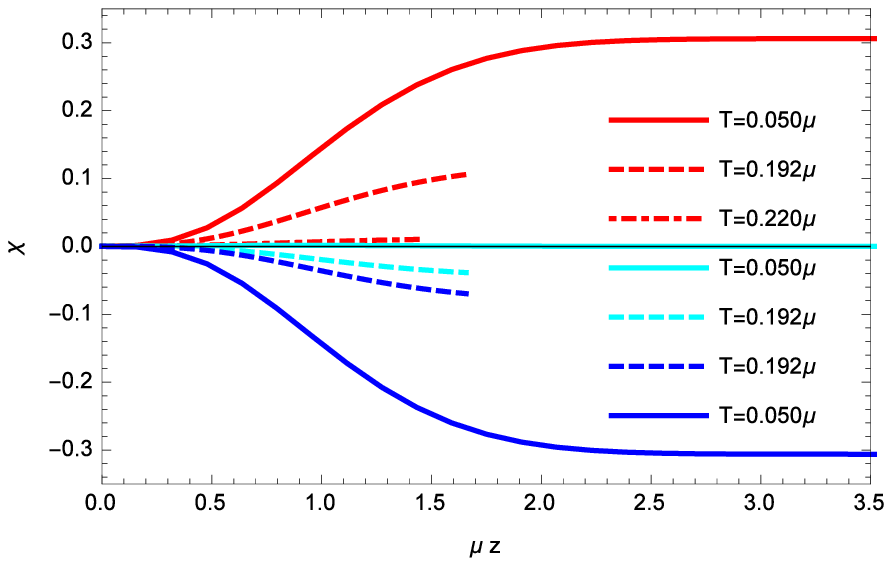} \vskip -0.05cm \hskip 0.15 cm
\textbf{( a ) } \hskip 6.5 cm \textbf{( b )} \\
\end{center}
\caption{Panel.(a) shows $\mu^3$ scaled dimensionless chiral condensate $\sigma(T)/\mu^3$ as a function of $\mu$ scaled temperature $T/\mu$ in negative quadratic dilaton background Eq.(\ref{negativedilaton}) when $v_3=0,v_4=8$ and $m_q=0.007\mu$.  Panel.(b) shows the corresponding $\chi$ solutions when $T=0.050,0.192,0.220\mu$. We only plot the part outside the black hole $0<z<z_h=1/(\pi T)$ in each temperature and cut the long constant tail in the region $3.5<\mu z<20/\pi$ when $T=0.05\mu$ for compactness of the plots.}
 \label{negative-sigma-m1}
\end{figure}

Now that the chiral phase transition in chiral limit is described quite well in this model, we would go further and try to investigate its finite quark mass behavior. We take $m_q=0.007\mu$ as an example and plot the results in Fig.\ref{negative-sigma-m1}. We would emphasize that the qualitative results for any finite quark mass are the same, though for simplicity we only show only one of them here. From Fig.\ref{negative-sigma-m1}(a), we see that the positive quark mass $m_q=0.007\mu$ would break the $\sigma\leftrightarrow-\sigma$ symmetry of the solution. As a result, the trivial $\chi\equiv0$ would not be a solution any more, since it does not satisfy the boundary condition $\chi(z)=m_q \zeta z+...$ at UV. Comparing to the chiral limit result Fig.\ref{negative-sigma-m0}, we could see that near $T=0$ the results do not change too much and the vacuum value of $\sigma$ is still around $0.08\mu^3$. However, when $T>0.15\mu$, the low $T$ cyan part and the high $T$ red part of the trivial solution $\chi\equiv0$ in Fig.\ref{negative-sigma-m0}(a) would be seperated when $m_q>0$ in Fig.\ref{negative-sigma-m1}(a). The cyan part is dragged down from the $x$ axis and bent towards the negative $\sigma$ branch, while the red part would be dragged up and bent towards the positive $\sigma$ branch. The blue non-trivial solution and the quark mass induced cyan part would join together at around $T=0.195\mu$, while the quark mass induced red part would join the red non-trivial solution and become a continuous line.

In Fig.\ref{negative-sigma-m1}(b), we also plot the solutions of $\chi(z)$ at different temperature $T=0.05,0.192,$ $0.22\mu$. From the figure, we could also see the same qualitative picture from the evolution of $\chi$. At low temperature like $T=0.05$, the three solutions do not change too much, and the cyan line almost equals to $0$. Then the increasing of temperature would drag the cyan lines down towards the blue negative $\sigma$ solutions, and it could be imagined that at temperature above $T=0.192\mu$ they would merge to be the same solution. Meanwhile, the red lines are dragged down towards the $x$ axis continuously.

\begin{figure}[h]
\begin{center}
\epsfxsize=7.5 cm \epsfysize=7.5 cm \epsfbox{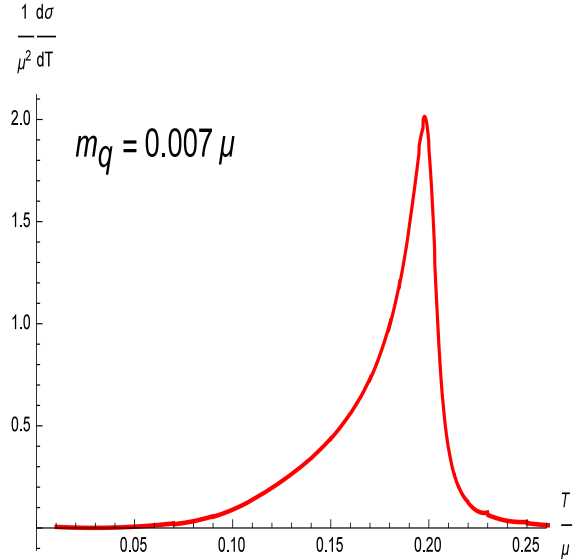} \
\end{center}
\caption{$\mu^2$ scaled $\frac{d \sigma(T)}{d T}$ as a function of $T$ in negative quadratic dilaton background Eq.(\ref{negativedilaton}) when $v_3=0,v_4=8$ and $m_q=0.007\mu$.} \label{dsigma-dT-m1-n}
\end{figure}

From the red dashed line in Fig.\ref{negative-sigma-m1}(a), we see that at finite quark mass the phase transition would become a crossover one, since the finite quark mass would destroy the $\sigma\leftrightarrow-\sigma$ symmetry in the 5D theory. The crossover transition is not a real phase transition, and we would define the transition point as the location of largest $\frac{d \sigma}{d T}$. We plot the results of $\frac{d \sigma}{d T}$ in Fig.\ref{dsigma-dT-m1-n}, and there we can see that since the quark mass we take are quite small, the transition temperature is almost the same as that in chiral limit, i.e.$T_C^{negative,m}\simeq0.198\mu$.

Thus, the negative dilaton plus the quartic scalar potential do give a very good realization of spontaneous chiral symmetry breaking in the vacuum and its restoration at high temperature. The transition is of second order type in chiral limit and it would turn to be a crossover one in any finite quark mass. However, as pointed out in \cite{Karch:2006pv,Li:2012ay,KKSS-2}, the large $z$ negative dilaton would cause an un-physical massless scalar meson state. Therefore, we will modify the large $z$ behavior of dilaton field and try to cure this problem in the next section.

\section{Interpolated dilaton: two different scales}
\label{sec-interpolation}

From the above discussion, we find that the negative dilaton gives a prediction on chiral phase transition agreeing perfectly well with the Columbia sketch plot whereas the positive dilaton background can not generate correct results in chiral limit. However, as pointed out in \cite{Karch:2006pv,Li:2012ay,KKSS-2}, the negative dilaton background predicts an un-physical massless scalar meson state which is unacceptable. The results from spectra analysis and thermodynamical analysis seem in contradiction. To solve this issue, one has to note that the dominating energy scale of chiral symmetry breaking is around $1\rm{GeV}$ and that of confinement is around $200\sim300\rm{MeV}$\cite{shuryakbook}. In the Regge behavior analysis, the confinement is the dominating effect while in chiral symmetry breaking mechanism new scale should be introduced. Taking into account the latter, we expect that the negative dilaton dominates at small $z$ and at large $z$ positive dilaton would dominate. In between the two, we take the following simple interpolation as a test
\begin{eqnarray}\label{int-dilaton}
\Phi(z)=-\mu_1^2z^2+(\mu_1^2+\mu_0^2)z^2\tanh(\mu_2^2z^2),
\end{eqnarray}
where $\mu_0$ would be fixed to around $0.43 {\rm GeV}$ as the extended soft-wall models \cite{Gherghetta-Kapusta-Kelley,Gherghetta-Kapusta-Kelley-2,YLWu,YLWu-1,Cui:2013xva,Li:2012ay,Li:2013oda,Colangelo:2008us} to produce
the correct Regge slope in the highly excited states, and $\mu_1, \mu_2$ are two free parameters. At UV, $\Phi(z)\rightarrow -\mu_1^2z^2$, and at IR the above interpolation form goes to $\Phi(z)\rightarrow\mu_0^2z^2$, which is responsible for the linear confinement. In this section, we will focus on solving the structure of chiral condensate under the dilaton profile Eq.(\ref{int-dilaton}) and the scalar potential Eq.(\ref{vchi}).

\subsection{$SU(2)$ case: $v_3=0, v_4\neq0$}
\label{sec-int-su2}

Since we fix $\mu_G=0.43\rm{GeV}$ for the Regge slope, there are 4 free parameters $\mu_1,\mu_2, v_3, v_4$ under the simple selection of dilaton profile Eq.(\ref{int-dilaton}) and the scalar potential Eq.(\ref{vchi}). As we mentioned in Sec.\ref{sec-quartic}, the t'Hooft determinant term would only appear in $SU(3)$ case, since $det[X]$ is a $\chi^2$ term in $SU(2)$ case. In this section, we would focus on $SU(2)$ case first, so we will take $v_3=0$. Thus, the only free parameters left are $\mu_1, \mu_2$ and $v_4$. In principle, we could fix this parameters by comparing the predicted meson spectral to the experimental data. However, considering that to get a better fitting of the experimental data quantitatively, the metric as well as the 5D $M_5^2$ could be modified as a more general $z$ dependent function like in \cite{Cui:2013xva,Cui:2014oba}. Here we would like to focus on the qualitative results and leave the careful quantitative study to the future. Since qualitatively, the model we are considering could well describe the linear confinement behavior at zero temperature, we would like to investigate the qualitative behavior of the chiral dynamics in this model.

\begin{figure}[h]
\begin{center}
\epsfxsize=6.5 cm \epsfysize=6.5 cm \epsfbox{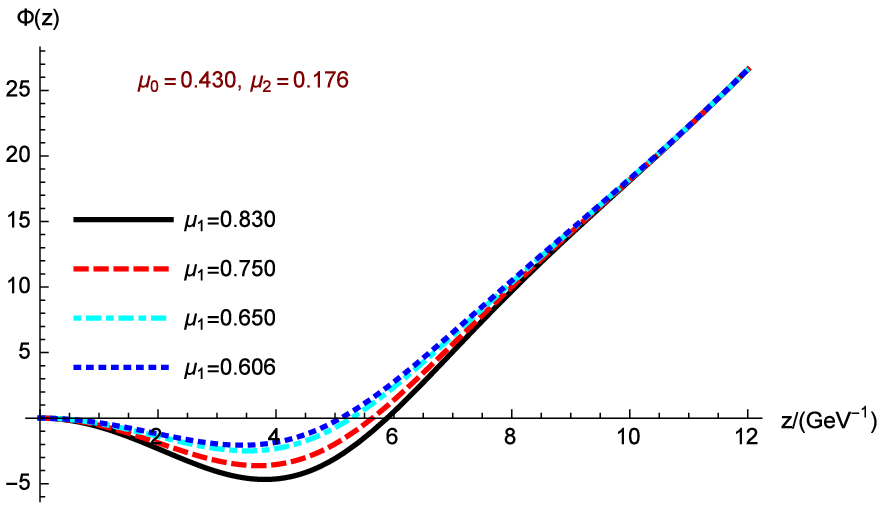}
\hspace*{0.1cm} \epsfxsize=6.5 cm \epsfysize=6.5 cm
\epsfbox{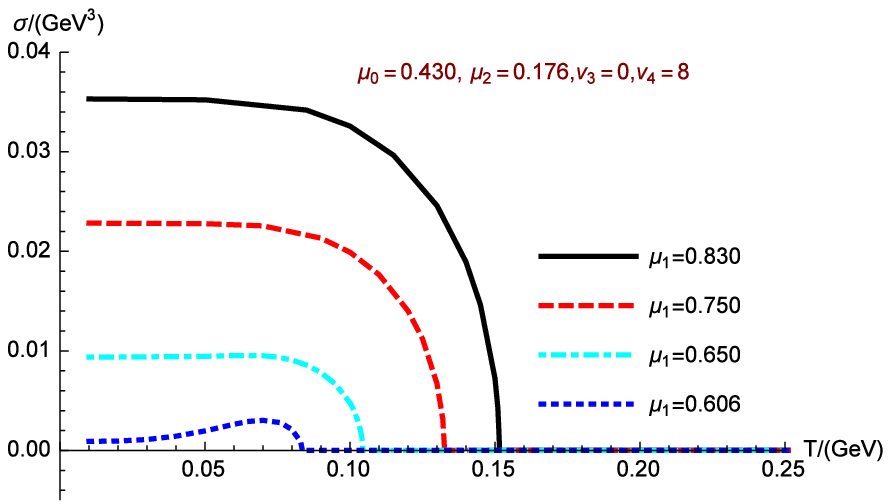} \vskip -0.05cm \hskip 0.15 cm
\textbf{( a ) } \hskip 6.5 cm \textbf{( b )} \\
\end{center}
\caption{$\mu_1$ dependence of $\Phi(z)$ in Eq.(\ref{int-dilaton}) and the corresponding results of $\sigma(T)$ as a function of temperature $T$ when $v_3=0,v_4=8$ and $\mu_0=0.43 \rm{GeV}, \mu_2=0.176\rm{GeV}$. In Panel.(a), the black solid, red dashed, cyan dotdashed, and blue dotted lines give the plots of $\Phi(z)$ when $\mu_1=0.83,0.75,0.65,0.606\rm{GeV}$ respectively. Here we only plot the $\sigma>0$ part, and the $\sigma<0$ part can be easily got by reflecting the results along $T$ axis. In Panel.(b) $\sigma(T)$ in chiral limit are shown in lines with the same symbols correspondingly.}
\label{sigma-int-mu1}
\end{figure}

Firstly, we would study the $\mu_1, \mu_2$ dependence of the model results in chiral limit. We take $v_4=8$ as in the negative dilaton case and $m_q=0$. As an example, we fix $\mu_2=0.176 \rm{GeV}$, then we tune $\mu_1$ and try to solve $\sigma(T)$ from Eq.(\ref{eom-chi-1}). We plot the results in Fig.\ref{sigma-int-mu1}. From Fig.\ref{sigma-int-mu1}(a), the negative part of dilaton field would be shift up when one decrease $\mu_1$, while the large $z$ quadratic tails stay in the same limit $\mu_0^2 z^2$. From Fig.\ref{sigma-int-mu1}(b), we can see that as in the negative dilaton model, the chiral condensate only appears in low temperature region. The transition temperature where the non-trivial solution start disappearing would increase with the increasing of $\mu_1$, i.e. the increasing of the negative area in dilaton profile. When $\mu_1=0.75, 0.83 \rm{GeV}$, chiral condensate would decrease monotonous from a finite value to zero, while for $\mu_1=0.606, 0.65\rm{GeV}$ chiral condensate would increasing with temperature first and then decrease rapidly to zero.

\begin{figure}[h]
\begin{center}
\epsfxsize=6.5 cm \epsfysize=6.5 cm \epsfbox{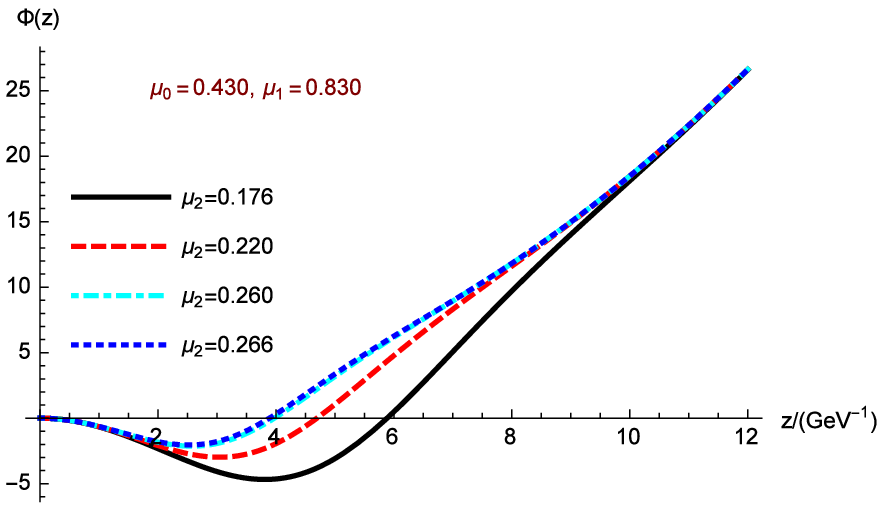}
\hspace*{0.1cm} \epsfxsize=6.5 cm \epsfysize=6.5 cm
\epsfbox{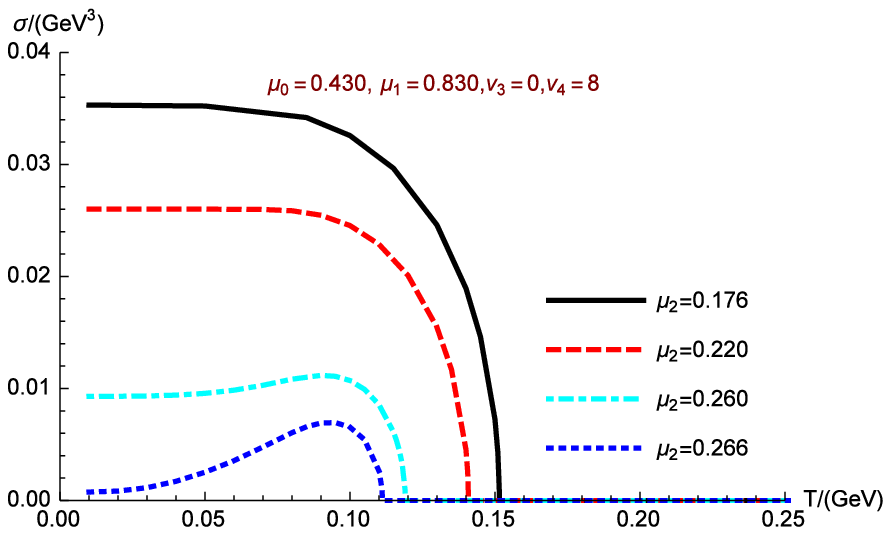} \vskip -0.05cm \hskip 0.15 cm
\textbf{( a ) } \hskip 6.5 cm \textbf{( b )} \\
\end{center}
\caption{$\mu_2$ dependence of $\Phi(z)$ in Eq.(\ref{int-dilaton}) and the corresponding results of $\sigma(T)$ as a function of temperature $T$ when $v_3=0,v_4=8$ and $\mu_0=0.43 \rm{GeV}, \mu_1=0.83\rm{GeV}$. In Panel.(a), the black solid, red dashed, cyan dotdashed, and blue dotted lines give the plots of $\Phi(z)$ when $\mu_1=0.176,0.22,0.26,0.266\rm{GeV}$ respectively. In Panel.(b) $\sigma(T)$ in chiral limit are shown in lines with the same symbols correspondingly.}
\label{sigma-int-mu2}
\end{figure}

Next, we fix $\mu_1=0.83\rm{GeV}$, take $v_4=8, m_q=0$ and turn to study the $\mu_2$ effects. The results are shown in Fig.\ref{sigma-int-mu2}. Similarly, from the figure we see that when we increase $\mu_2$ the negative part of dilaton would be depressed and the transition temperature and chiral condensate would decrease. What's more, when the negative part of dilaton profile become small, chiral condensate start to increase at low temperature region and decrease fast to zero near the transition temperature.

From the above dependence behavior of $\mu_1, \mu_2$, we could see that a larger negative part of dilaton would favor a larger $\sigma$ in the vacuum and low temperature region. This is consistent with our previous study on the positive and negative quadratic models, in which no negative part of dilaton field is corresponding to zero chiral condensate and full negative part gives finite condensate in chiral limit. In addition, this also shows that it is the intermediate part of the dilaton dominating in chiral dynamics, since the large $z$ behavior of the model is universal for different parameter values.

\begin{figure}[h]
\begin{center}
\epsfxsize=6.5 cm \epsfysize=6.5 cm \epsfbox{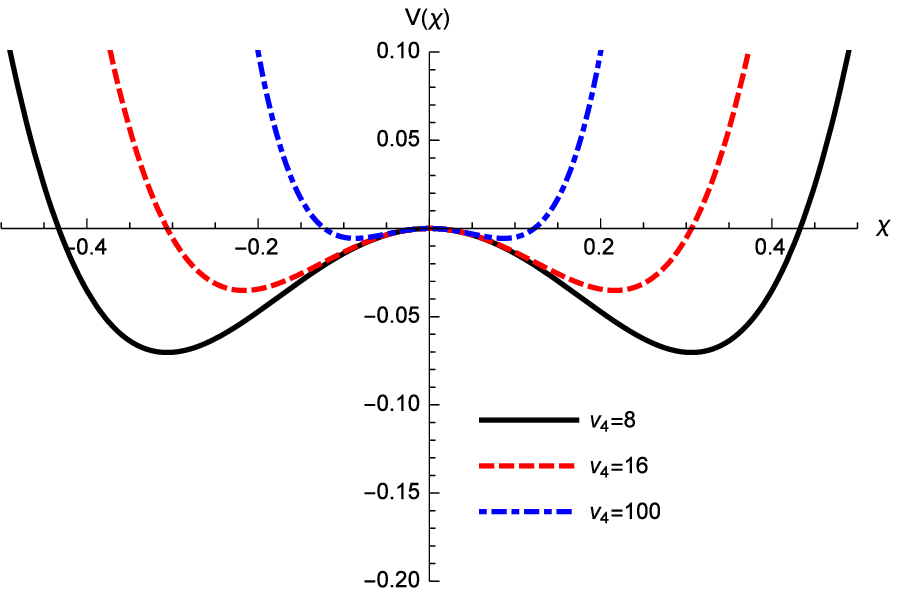}
\hspace*{0.1cm} \epsfxsize=6.5 cm \epsfysize=6.5 cm
\epsfbox{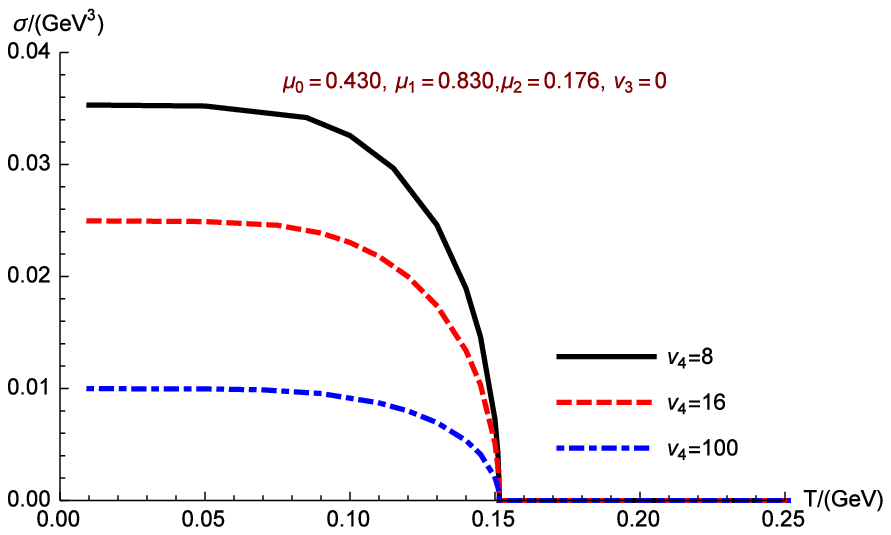} \vskip -0.05cm \hskip 0.15 cm
\textbf{( a ) } \hskip 6.5 cm \textbf{( b )} \\
\end{center}
\caption{$v_4$ dependence of the scalar potential $V(\chi)$ in Eq.(\ref{vchi}) and the corresponding results of $\sigma(T)$ as a function of temperature $T$ when $v_3=0$ and $\mu_0=0.43 \rm{GeV}, \mu_1=0.83\rm{GeV}, \mu_2=0.176\rm{GeV}$. In Panel.(a), the black solid, red dashed and blue dotdashed lines give the plots of $V(\chi)$ when $v_4=8,16,100$ respectively. In Panel.(b) $\sigma(T)$ in chiral limit are shown in lines with the same symbols correspondingly.}
\label{sigma-int-v4}
\end{figure}

Finally, we take $\mu_1=0.83\rm{GeV},\mu_2=0.176\rm{GeV}$ and study the model dependence of $v_4$. The results are presented in Fig.\ref{sigma-int-v4}. From Fig.\ref{sigma-int-v4}(a), we can see that increasing $v_4$ would decrease the non-trivial vacuum $\chi_0=\pm\sqrt{\frac{3}{4v_4}}$. Correspondingly, the chiral condensate would decrease, as can be seen from Fig.\ref{sigma-int-v4}(b). However, unlike decreasing $\mu_1$ and increasing $\mu_2$, the transition temperature would stay the same when increase $v_4$.

\begin{figure}[h]
\begin{center}
\epsfxsize=6.5 cm \epsfysize=6.5 cm \epsfbox{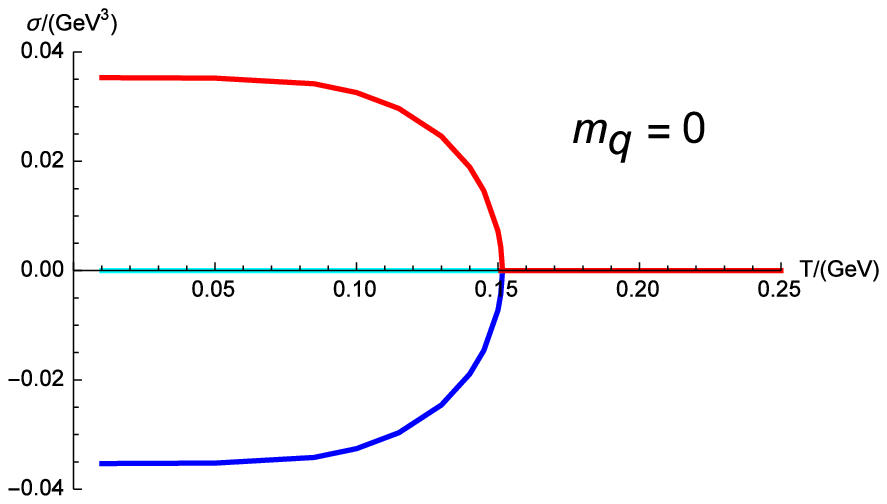}
\hspace*{0.1cm} \epsfxsize=6.5 cm \epsfysize=6.5 cm
\epsfbox{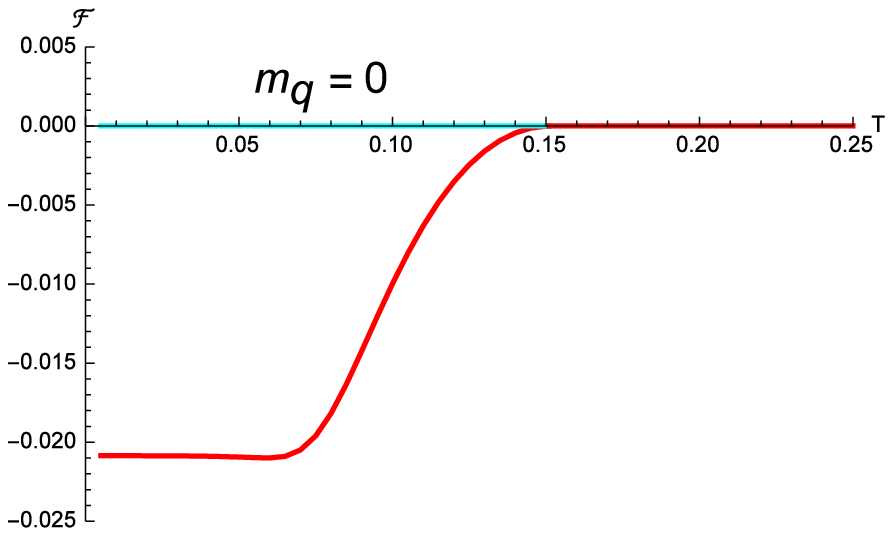} \vskip -0.05cm \hskip 0.15 cm
\textbf{( a ) } \hskip 6.5 cm \textbf{( b )} \\
\end{center}
\caption{The results of chiral condensate $\sigma(T)$ and the corresponding free energy density $\mathcal{F}$ as a function of temperature $T$ in dilaton background Eq.(\ref{int-dilaton}) when $v_3=0,v_4=8$ and $\mu_0=0.43\rm{GeV}, \mu_1=0.83\rm{GeV},\mu_2=0.176\rm{GeV}$. Panel.(a) shows the result of $\sigma(T)$ in chiral limit and Panel.(b) compares the free energy density $\mathcal{F}$ of different solutions.}
\label{interpolation-sigma}
\end{figure}

Based on the above studies, we take $\mu_1=0.83\rm{GeV},\mu_2=0.176\rm{GeV}, v_4=8$ and turn to study the effects of quark mass. As before, we plot the results of chiral limit in Fig.\ref{interpolation-sigma} first. From Fig.\ref{interpolation-sigma}(a), we can read that the vacuum value of chiral condensate $\sigma_0$ is around $0.035\rm{GeV}^3\simeq(327\rm{MeV})^3$ and the transition temperature $T_C^{SU(2),0}$ is around $151\rm{MeV}$. Both of them are comparable with the lattice results. Furthermore, since in chiral limit the free energy formula Eq.(\ref{freeenergy}) is finite near the boundary $z=0$, we insert the $\chi(z)$ solutions into this formula and get the free energy results as shown in Fig.(\ref{interpolation-sigma})(b). From this figure, we could see that the free energy of the non-trivial solutions are always smaller than the trivial solutions. Besides, near the transition point, we find that the curve of non-trivial solutions is tangent to the trivial solution curve. This fact confirm that the phase transition in chiral limit is a second order one.

\begin{figure}[h]
\begin{center}
\epsfxsize=6.5 cm \epsfysize=6.5 cm \epsfbox{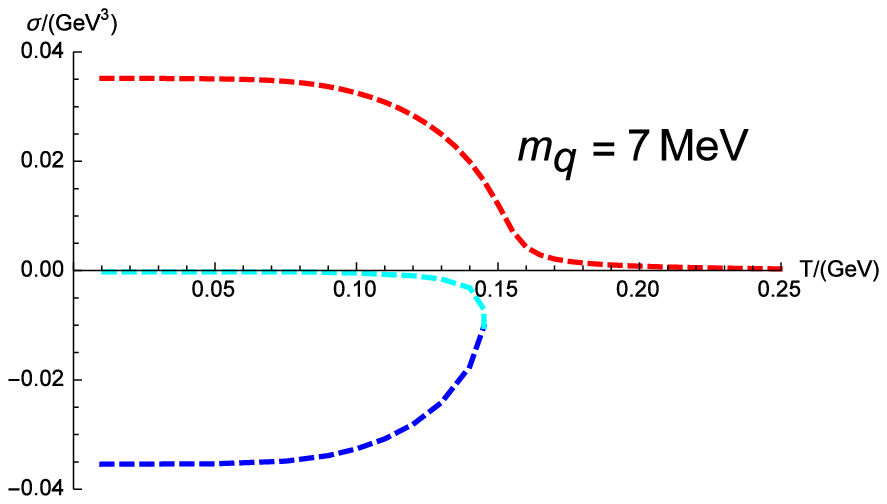}
\hspace*{0.1cm} \epsfxsize=6.5 cm \epsfysize=6.5 cm
\epsfbox{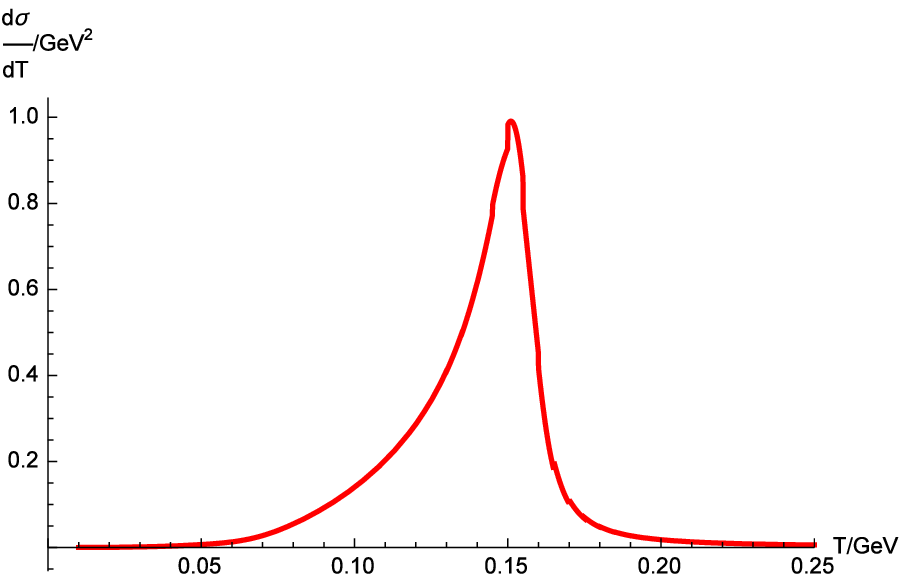} \vskip -0.05cm \hskip 0.15 cm
\textbf{( a ) } \hskip 6.5 cm \textbf{( b )} \\
\end{center}
\caption{The results of chiral condensate $\sigma(T)$ and its temperature derivative $\frac{d\sigma(T)}{d T}$ as a function of temperature $T$ in dilaton background Eq.(\ref{int-dilaton}) when $v_3=0,v_4=8$ and $\mu_0=0.43\rm{GeV}, \mu_1=0.83\rm{GeV},\mu_2=0.176\rm{GeV}$. Panel.(a) shows the results of $\sigma(T)$ when $m_q\simeq 7 MeV$. Panel.(b) gives the plot of $\frac{d\sigma(T)}{d T}$, the local maximum of which gives the pesudo transition temperature.}
\label{interpolation-sigma-1}
\end{figure}

After turning on the mass source, we find that any finite quark mass would change the phase transition order. As an explicit example, we take $m_q=7\rm{MeV}$ and show the results in Fig.\ref{interpolation-sigma-1}. As can be seen from Fig.\ref{interpolation-sigma-1}, the qualitative results are totally the same as in the negative dilaton model described in Sec.\ref{sec-quartic-negative}. A positive quark mass would shift up the high temperature trivial solutions, join the positive branches of non-trivial solutions and merge to be one continuous line. The low temperature trivial solutions would be shift down. It would join the negative non-trivial solutions and merge to be a semi-circle appearing only at temperature below $147\rm{MeV}$. Therefore, the physical line should be the continuous red line in Fig.\ref{interpolation-sigma-1}. To extract the pseudo-transition temperature, we calculate $\frac{d\sigma}{dT}$ and read the transition temperature from the peak of Fig.\ref{interpolation-sigma-1}(b), which gives $T_C^{SU(2),m}\simeq 150\rm{GeV}$, almost the same as the chiral limit.

As a short summary of this section, the interpolating dilaton model does give correct behavior of chiral phase transition. In chiral limit, it does give spontaneous chiral symmetry breaking in the vacuum and its restoration at high temperature. We also confirm that the transition order is of second order. At any finite quark mass the transition turns to be a crossover one. From the current study, the negative part in a intermediate scale is important for spontaneous chiral symmetry breaking and the positive quadratic part is important for linear confinement. Qualitatively, this is consistent with the studies in \cite{shuryakbook}, though we do not know how to map the exact value of the energy scales. It should also be noted that, here we choose a simple interpolating form of the dilaton field, which could be carefully tuned to accommodate the meson spectra also. In this work, we focus only on producing the correct qualitative behavior of chiral phase transition.

\subsection{$SU(3)$ case: $v_3,v_4\neq0$}
\label{sec-int-su3}

In the previous section, we have realized the top line of the phase diagram in mass plane, which is shown in Fig.\ref{columbia-plot}. In this section, we would like to test this model in the $SU(3)$ case, i.e. the diagonal line in Fig.\ref{columbia-plot}.

In the $SU(3)$ case, if we only consider the quartic term $X^4$, then it is easy to understand that the results would not be different from the $SU(2)$ case. Fortunately, it is possible to introduce the t'Hooft determinant term $\text{det}[X]$. If we consider a more general $N_f=2+1$ case, the vacuum expectation value of $X$ should be written as $X=\text{diag}\{\chi_l,\chi_l,\chi_s\}$. Then the mass term and the quartic term would be proportional to $2\chi_l^2+\chi_s^2$ and $2\chi_l^4+\chi_s^4$ respectively. There is no coupling term in between $\chi_l$ and $\chi_s$. It is different for the determinant term, which reduces to $\chi_l^2\chi_s$ and cause the mixing of light flavor and heavy one. In principle, we can study the whole plane in Fig.\ref{columbia-plot}. However, it is much more complicated to solve the two coupled equations. So, we would leave this calculation to the future and focus on the $SU(3)$ case only. When $m_u=m_d=m_s$, one can expect that $\chi_l=\chi_s$, then the above settings would reduce to the model described in previous section with finite $v_3$.

\begin{figure}[h]
\begin{center}
\epsfxsize=7.5 cm \epsfysize=7.5 cm \epsfbox{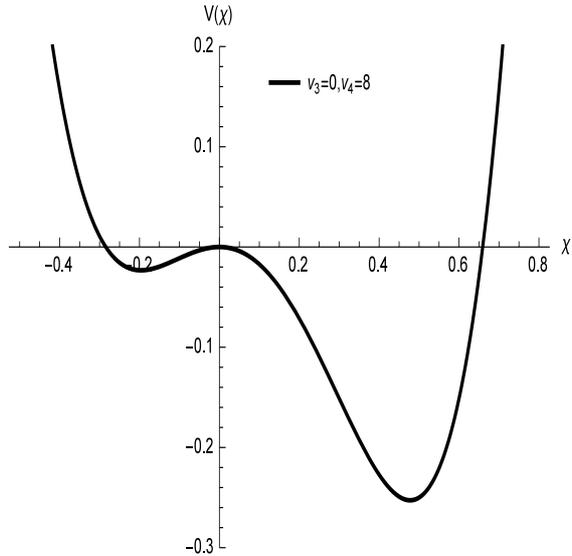} \
\end{center}
\caption{$V(\chi)$ as a function of $\chi$ with $v_3=-3,v_4=8$.} \label{v4=8v3=-3-potential}
\end{figure}

Since we are more interested in the qualitative result, instead of fixing $v_3, v_4$ from meson spectra, we would only take $v_3=-3, v_4=8$ as an example to show the qualitative behavior. The potential are shown in Fig.\ref{v4=8v3=-3-potential}. In the figure, we find that the $\chi\leftrightarrow-\chi$ symmetry is broken explicitly by the cubic potential. The right vacuum is around $\chi^{(+)}=0.477$ and the left vacuum is around $\chi^{(-)}=-0.196$. From the potential level, we could see that the right vacuum is more stable due to the lower value of potential energy.

\begin{figure}[h]
\begin{center}
\epsfxsize=6.5 cm \epsfysize=6.5 cm \epsfbox{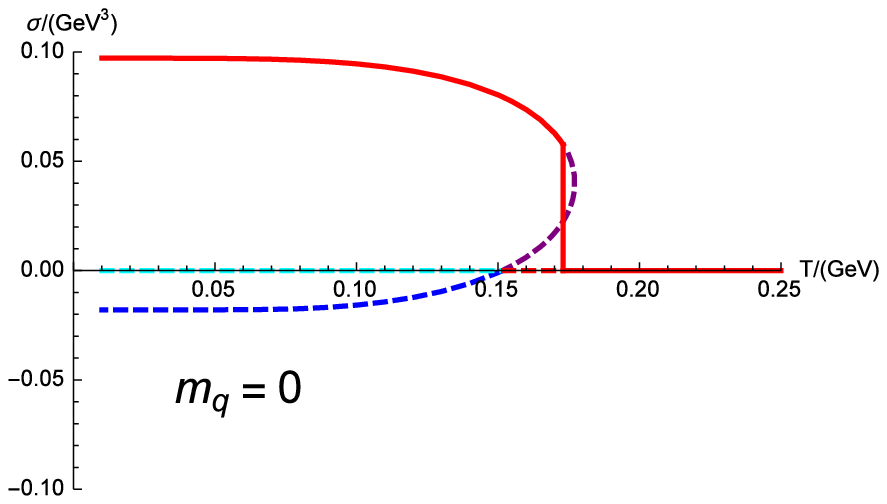}
\hspace*{0.1cm} \epsfxsize=6.5 cm \epsfysize=6.5 cm
\epsfbox{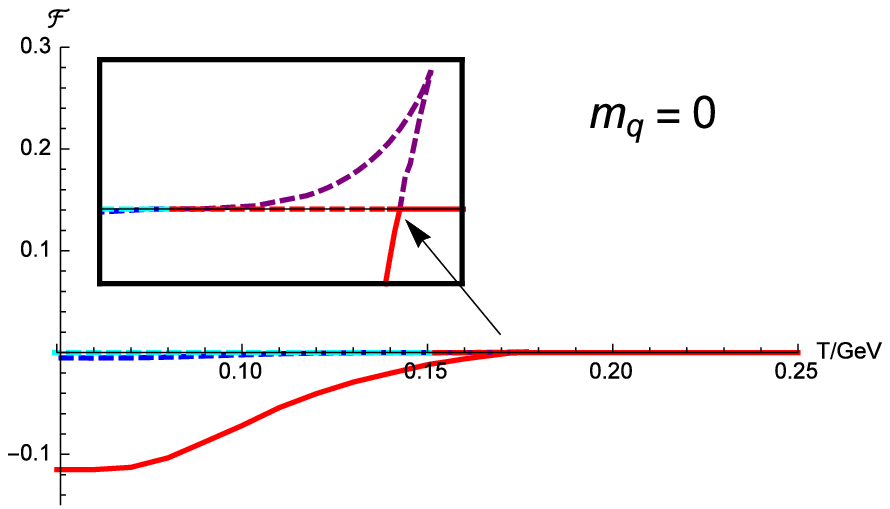} \vskip -0.05cm \hskip 0.15 cm
\textbf{( a ) } \hskip 6.5 cm \textbf{( b )} \\
\end{center}
\caption{The results of chiral condensate $\sigma(T)$ in chiral limit and the corresponding free energy density $\mathcal{F}$ as a function of temperature $T$ in dilaton background Eq.(\ref{int-dilaton}) when $v_3=-3,v_4=8$ and $\mu_0=0.43\rm{GeV}, \mu_1=0.83\rm{GeV},\mu_2=0.176\rm{GeV}$. Panel.(a) shows the result of $\sigma(T)$ in chiral limit. Panel.(b) compares the free energy density $\mathcal{F}$ of different solutions. The points representing the same solutions are labeled with the same colors and styles in Panel.(a) and Panel.(b).}
 \label{interpolation-sigma-v3}
\end{figure}

Inserting the potential into the equation of motion Eq.(\ref{eom-chi-1}), we solved the chiral condensate in chiral limit as shown in Fig.\ref{interpolation-sigma-v3}. From Fig.\ref{interpolation-sigma-v3}(a), we find that at low temperature there are still two branches of non-trivial solutions together with the trivial $\chi\equiv0$ solution, while above $T=0.151\rm{MeV}$ the non-trivial solutions disappears.  Due to the cubic term, we could see that the positive(the red solid line plus the purple dashed line) and negative non-trivial branches(the blue dashed line) are no longer $\sigma\leftrightarrow-\sigma$ reflection of each other. As we read from Fig.\ref{v4=8v3=-3-potential}, the potential energy of the right vacuum is lower. Therefore, we might expect that the positive sigma solutions would be more stable even in the chiral limit. To confirm this, we insert the $\chi(z)$ solution to the free energy formula Eq.\ref{freeenergy} and calculate the free energy density. The results are plotted in Fig.\ref{interpolation-sigma-v3}(b). From this plot, we do find that the free energy of the positive branch is smaller than the negative branch. As a result, the lower temperature region is dominated by the right vacuum. Then from the zooming out zone of Fig.\ref{interpolation-sigma-v3}(b), we find that the free energy density of the positive branch would start to be larger than zero, which shows that a phase transition would happen. The real physical path when increasing temperature would be the red solid line. The chiral condensate decrease from the finite vacuum value monotonously and at around $T=173\rm{MeV}$ it suddenly drop to zero, which gives a first order picture of the chiral restoration phase transition. This chiral limit result is just the same as expected from the diagonal line in Fig.\ref{columbia-plot}.

\begin{figure}[h]
\begin{center}
\epsfxsize=9.5 cm \epsfysize=7.5 cm \epsfbox{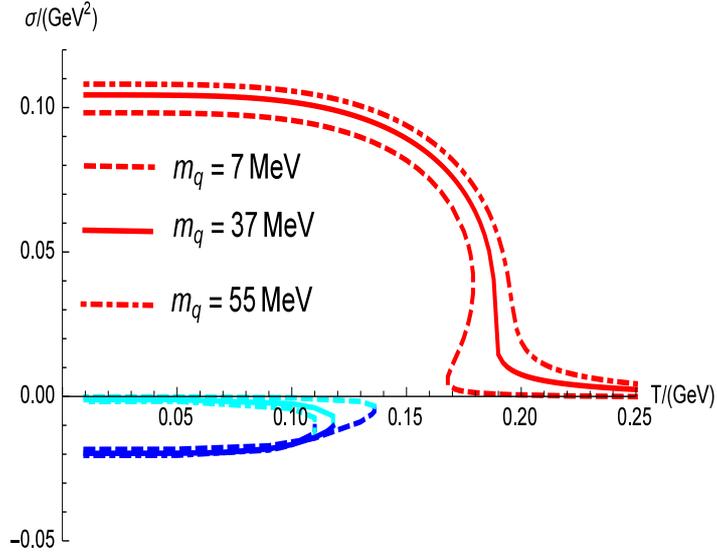} \
\end{center}
\caption{The results of chiral condensate $\sigma(T)$ as a function of temperature $T$ in dilaton background Eq.(\ref{int-dilaton}) when $v_3=-3,v_4=8$ and $\mu_0=0.43\rm{GeV}, \mu_1=0.83\rm{GeV},\mu_2=0.176\rm{GeV}$.  The red, cyan and blue dashed lines give the results of $m_q=7\rm{MeV}$ and the solid and dotdashed lines give the results of $m_q=37\rm{MeV}, m_q=55\rm{MeV}$ respectively.}
\label{interpolation-sigma-v3-mass}
\end{figure}

Now we turn the quark mass on. We find the low temperature $\chi\equiv0$ would join the negative non-trivial branch to form a semi-circle again, while the high temperature $\chi\equiv0$ branch would join the positive non-trivial branch and form a continuous line extend to high temperature. However, when $m_q$ is small, for example $m_q=7\rm{MeV}$ as shown in the dashed line of Fig.\ref{interpolation-sigma-v3-mass}, the positive branch would be multivalued in the temperature range $167\rm{MeV}<T<175\rm{MeV}$, which implies the first order property of the phase transition.  In addition, when $m_q$ is large enough, for example $m_q=55\rm{MeV}$ as shown in the dotdashed line in Fig.\ref{interpolation-sigma-v3-mass}, the positive branch would decrease monotonously and continuously, which implies a crossover transition. In between these two types, it is not difficult to imagine that with certain value of $m_q$ the derivative $\frac{d\sigma}{dT}$ would be divergent at certain temperature, just as shown in the solid line in Fig.\ref{interpolation-sigma-v3-mass}(around $T=195\rm{MeV}$). This mass would be the critical mass, which gives a second order phase transition. With the current parameter, we find that the critical mass is around $m_q=37\rm{MeV}$.

\begin{figure}[h]
\begin{center}
\epsfxsize=9.5 cm \epsfysize=7.5 cm \epsfbox{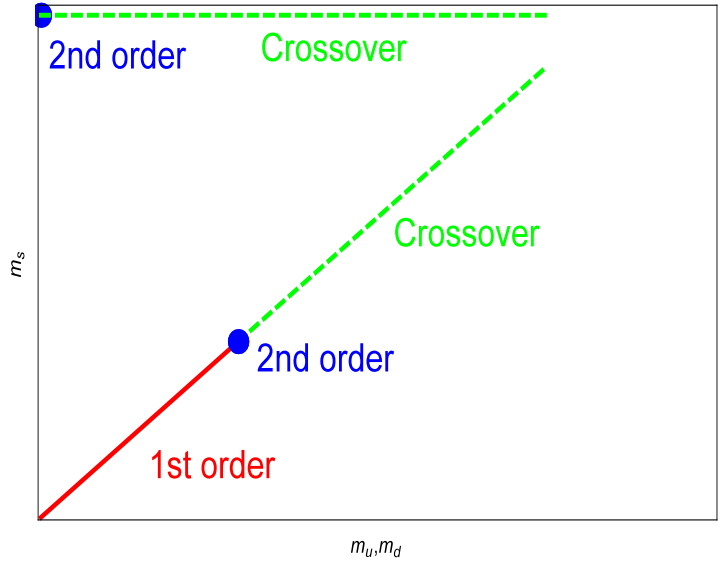} \
\end{center}
\caption{The $SU(2)$ and $SU(3)$ mass diagram in dilaton background Eq.(\ref{int-dilaton}) when $v_3=-3,v_4=8$ and $\mu_0=0.43\rm{GeV}, \mu_1=0.83\rm{GeV},\mu_2=0.176\rm{GeV}$.  The red solid line represents first order phase transition and the green dashed lines stand for crossover. The blue dots are second order points.}
\label{mass-diagram}
\end{figure}

In a short summary, when we turn on a negative cubic scalar potential, the model Eq.(\ref{int-dilaton}) gives a first order chiral phase transition in chiral limit and in small quark mass region, while in sufficient large quark mass area, the phase transition would turn to be a crossover one. We summarize the results for $SU(2)$ and $SU(3)$ in Fig.\ref{mass-diagram}. This agrees well with the picture shown in Fig.\ref{columbia-plot}(the top line and the diagonal line). Combining the results in $SU(2)$ case and $SU(3)$ case, we might expect that when extend to $N_f=2+1$ case, the model could produce phase diagram in mass plane similar to Fig.\ref{columbia-plot} qualitatively. We will leave this calculation in the future.

\section{Conclusion and discussion}
\label{sum}
In this work, we investigate the chiral phase transition within the soft-wall holographic framework. By imposing proper UV and IR boundary conditions, we could solve chiral condensate as a function of quark mass and temperature. We first try in the original soft-wall model with a pure positive quadratic dilaton and a mass term in the scalar potential. The results shown that in chiral limit there is no chiral condensate in the original soft-wall model, which implies that the chiral condensate in the original soft-wall model is induced by quark mass and the symmetry breaking is an explicit one. After introducing a quartic term in scalar potential to break the linear property of the equation of motion, we still find that in chiral limit the pure quadratic dilaton model can not realize the spontaneous chiral symmetry breaking.

Then, we consider the pure negative quadratic dilaton case with a positive quartic potential. We show that in this model, both the spontaneous chiral symmetry breaking in the vacuum and its restoration are realized correctly. The phase transition is a second order one in chiral limit and it turns to be a crossover one at any finite quark mass.

However, the negative part of dilaton field at large $z$ region would cause an massless scalar meson state, which has never been detected experimentally and is unacceptable. To cure this problem, we propose a dilaton model, which tends to be negative quadratic at small $z$ region and positive quadratic at large $z$ region. The intermediate interpolating configuration is controlled by two mass parameters. We find that in a large parameter region, the model can realize chiral phase transition well. In $SU(2)$ case with positive value of the quartic scalar coefficient, it gives a second order phase transition in chiral limit and a crossover one at any finite quark mass. In $SU(3)$ case with a negative cubic and positive quartic potential term, it gives a first order phase transition in chiral limit and small quark mass region, while in sufficient quark mass case, the phase transition turns out to be crossover again. Qualitatively, the results agrees perfectly with the Colombia sketch plot as shown in Fig.\ref{columbia-plot}. Quantitatively, we also show that if one take the proper value of the parameters, then the vacuum value of $\sigma$ would be around $(320\rm{MeV})^3$ and the transition temperature is around $150\rm{MeV}$, which are consistent with the lattice results. The studies here also show that the dominant scales for chiral dynamics and confinement are different, which is consistent with the previous studies in \cite{shuryakbook}.

Finally, we would like to emphasize that we have not considered the meson spectra in a quantitative way. Instead, we only focus on the linear confinement, i.e. the Regge behavior in highly excited state. To generate meson spectra comparable to the experiment data, the simple interpolation selected in this work might be over simplified. We might need to modify the IR behavior of several quantities together. In general, such an IR modification could be imposing on metric, quartic term, dilaton and conformal mass as in \cite{Cui:2013xva,Cui:2014oba} to accommodate meson spectra simultaneously. Furthermore, we expect when extending to $N_f=2+1$ case, our model could generate the same qualitative results as shown in Fig.\ref{columbia-plot}. We will leave these work to the future.

\vskip 0.5cm
{\bf Acknowledgement}
\vskip 0.2cm
The authors thank Song He, Yi Yang for valuable discussions. K.C is supported by CAS-TWAS president fellowship.
M.H. is supported by the NSFC under Grant Nos. 11175251 and 11275213, DFG and NSFC (CRC 110),
CAS key project KJCX2-EW-N01, and Youth Innovation Promotion Association of CAS. This work is funded in part by China Postdoctoral Science Foundation.

\appendix

\section{Numerical Method}
\label{appendix-sec1}
In this section, we will describe how to determine chiral condensate numerically in the soft-wall model. Without loss of generality, we will take the simple negative dilaton $\Phi(z)=-\mu^2z^2$ in the quartic potential model as an example. In this case, Eq.(\ref{eom-chi-1}) becomes

\begin{eqnarray}\label{app-eom}
\chi^{''}-(\frac{3}{z}-2\mu^2 z+\frac{4\pi^4T^4 z^3}{1-\pi^4T^4z^4})\chi^{'}+ \frac{1}{z^2(1-\pi^4T^4z^4)}(3\chi-4v_4 \chi^3)=0.
\end{eqnarray}

In principle, the above second order ordinary differential equation has two independent solutions. However, as easily can be read, there is an apparent singular point $z=z_h=1/(\pi T)$ in the equation. This fact would lead one of the two independent solutions singular at this point. Therefore, if one requires the physical solution to be regular at the black hole horizon, then we can determine one of the two boundary coefficients from the other.

\begin{figure}[h]
\begin{center}
\epsfxsize=6.5 cm \epsfysize=6.5 cm \epsfbox{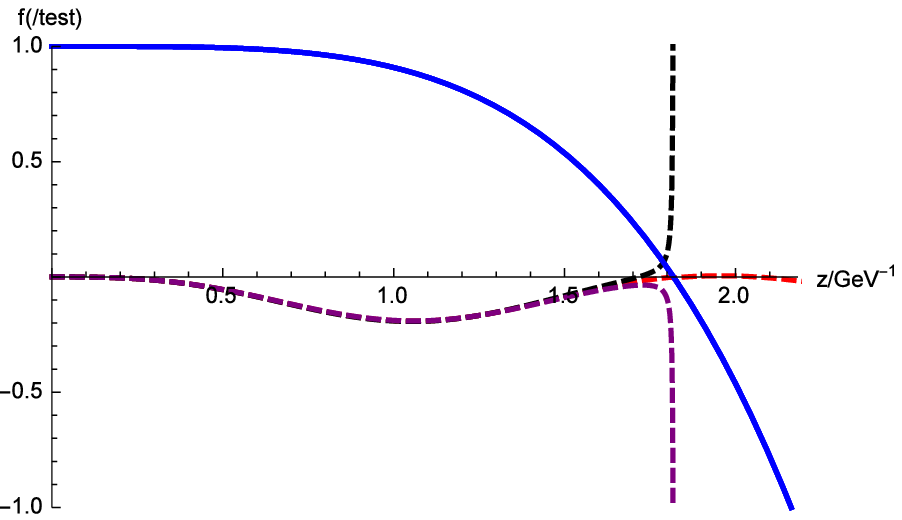}
\hspace*{0.1cm} \epsfxsize=6.5 cm \epsfysize=6.5 cm
\epsfbox{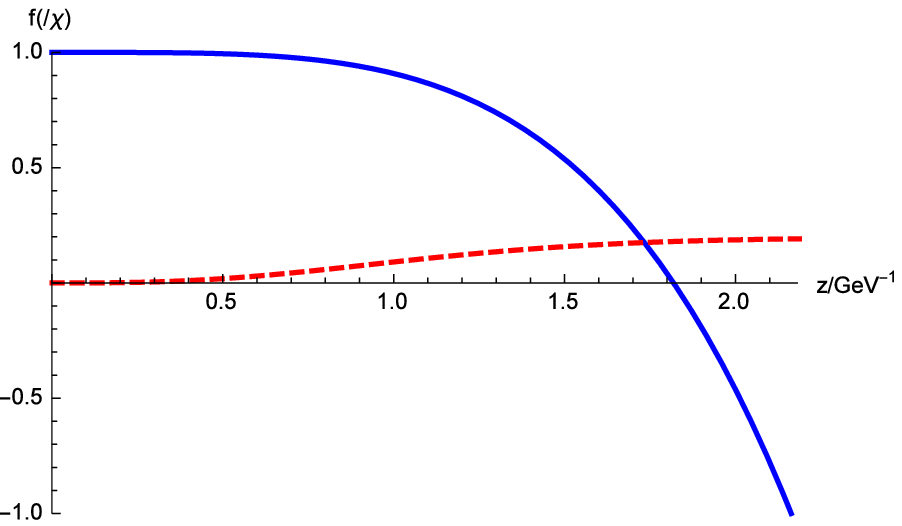} \vskip -0.05cm \hskip 0.15 cm
\textbf{( a ) } \hskip 6.5 cm \textbf{( b )} \\
\end{center}
\caption{$test(z)$ as a function of $z$ in negative quadratic dilaton background Eq.(\ref{negativedilaton}) when $v_3=0,v_4=8$ and $m_q=0,\mu=1 \rm{GeV}, T=0.175\rm{GeV}$. Panel.(a): the blue solid line represents $f(z)$ and the black, red, purple dashed lines represent for $test(z)$ solutions when $\sigma=0.050,0.0494742...,0.049$ respectively. Panel.(b): the blue solid line represents $f(z)$ and the red dashed line represents $\chi(z)$ when $\sigma=0.0494742$.}
 \label{test-f-chi}
\end{figure}

As mentioned in the previous sections, the boundary expansion of Eq.(\ref{app-eom}) is of the following form

\begin{eqnarray}
\chi(z)=m_q \zeta z+\frac{\sigma}{\zeta} z^3+...,
\end{eqnarray}
where $m_q$ is mapped to the current quark mass and $\sigma$ is mapped to chiral condensate. As an example, here we take $m_q=0$. Isolating the apparent singular part of second order eqaution, we get $\frac{f^{'}\chi^{'}-e^{2A_s}\partial_\chi V(\chi)}{f}$. Noting that to get a regular solution is equivalent to get a solution satisfying $\frac{f^{'}\chi^{'}-e^{2A_s}\partial_\chi V(\chi)}{f}=0$, one can take the test function $test\equiv -z^2(\frac{f^{'}\chi^{'}-e^{2A_s}\partial_\chi V(\chi)}{f})$ as a signal and tune $\sigma$ towards the correct value. In Fig.\ref{test-f-chi}(a), we take $T=0.175\rm{GeV}$ as an example and show this process explicitly. At the beginning, when we take $\sigma=0.050$, we see that at the horizon the test function blows up towards positive infinity(see the black dashed line). Then, we decrease $\sigma$ to $0.049$, and find that the test function blows up towards negative infinity (see the purple dashed line). This means that the physical value of $\sigma$ is between $0.050$ and $0.049$. Therefore, we then try $\sigma$ to be the middle value of $0.050$ and $0.049$. Repeating the process for several times, we find that when $\sigma=0.0494742...$, the test function can go smoothly through horizon(see the red dashed line), and we take this value as the physical value. In Fig.\ref{test-f-chi}(b), we show that when $\sigma=0.0494742$, the numerical solution of $\chi$ can go smoothly through the horizon also.

\end{document}